\documentclass[journal]{IEEEtran}

 \usepackage{graphicx} 

  \DeclareGraphicsExtensions{.eps}
  
  \usepackage{subcaption} 
  
  \usepackage{tikz}
  \usepackage{pgfplots,pgfplotstable}
  \usetikzlibrary{pgfplots.groupplots}
  \pgfplotsset{compat=1.5}

\usepackage{multirow}
\usepackage{colortbl} 

\usepackage{amsmath}

\usepackage{alphalph,etoolbox}

\usepackage{siunitx}

\AtBeginDocument{%
  \AtBeginEnvironment{subequations}{%
  }
}

\usepackage{textcomp}
\usetikzlibrary{shapes,arrows}

\usepackage{empheq} 

\usepackage{stfloats} 
\usepackage[hyphens]{url}
\usepackage[hidelinks]{hyperref}
\hypersetup{colorlinks=true,linkcolor=black,citecolor=black,filecolor=black,urlcolor=black,breaklinks=true}
\usepackage{cite} 

\usepackage{color}

\newcommand{\norm}[1]{\left\lVert#1\right\rVert} 

\usepackage{cancel} 

\usepackage{array} 

\usepackage[export]{adjustbox}

\usepackage[normalem]{ulem}

\usepackage{amsfonts} 

\usepackage{diffcoeff}

\usepackage{verbatim}
\usepackage{forest}
\usetikzlibrary{arrows.meta, shapes.geometric, calc, shadows,fit}

\colorlet{mygreen}{green!75!black}
\colorlet{col1in}{red!30}
\colorlet{col1out}{red!40}
\colorlet{col2in}{mygreen!40}
\colorlet{col2out}{mygreen!50}
\colorlet{col3in}{blue!30}
\colorlet{col3out}{blue!40}
\colorlet{col4in}{mygreen!20}
\colorlet{col4out}{mygreen!30}
\colorlet{col5in}{blue!10}
\colorlet{col5out}{blue!20}
\colorlet{col6in}{blue!20}
\colorlet{col6out}{blue!30}
\colorlet{col7out}{orange}
\colorlet{col7in}{orange!50}
\colorlet{col8out}{orange!40}
\colorlet{col8in}{orange!20}
\colorlet{linecol}{blue!60}

\makeatletter
\def\bstctlcite{\@ifnextchar[{\@bstctlcite}{\@bstctlcite[@auxout]}}
\def\@bstctlcite[#1]#2{\@bsphack
 \@for\@citeb:=#2\do{%
   \edef\@citeb{\expandafter\@firstofone\@citeb}%
   \if@filesw\immediate\write\csname #1\endcsname{\string\citation{\@citeb}}\fi}%
 \@esphack}
\makeatother

\begin{document}

\title{Strategic Bidding in the Frequency-Containment Ancillary Services Market}

\author{Carlos~Matamala
        and~Goran~Strbac

\vspace{-8mm}
}

\markboth{}
{Shell \MakeLowercase{\textit{et al.}}: Bare Demo of IEEEtran.cls for IEEE Journals}

\maketitle
\begin{abstract}
The vast integration of non-synchronous renewable energy sources compromises power system stability, increasing vulnerability to frequency deviations due to the lack of inertia. Current efforts to decarbonise electricity grids while maintaining frequency security still rely on Ancillary Services (AS) provision, such as inertia and frequency response, from flexible synchronous generators, placing these type of units in an advantageous position in the AS market. However, in the ongoing transition to decarbonisation, not enough attention has been given to analysing market power in the frequency-containment AS market.

This work presents a strategic bidding model designed to analyse market power in the coupled energy and frequency-containment AS market. Through a non-convex primal-dual bi-level formulation, we determine the interaction of a strategic market player with the rest of the market that behaves competitively. The case study is based on Great Britain in 2030, demonstrating the capacity of the strategic player to influence prices. While this impact is perceived in the energy market, it is particularly pronounced in the AS market.
\end{abstract}

\vspace*{-1mm}

\IEEEpeerreviewmaketitle

\vspace*{-3mm}
\section*{Nomenclature}
\addcontentsline{toc}{section}{Nomenclature}

\vspace{-3pt}

\subsection*{Indices and Sets}
\begin{IEEEdescription}[\IEEEusemathlabelsep\IEEEsetlabelwidth{$P_{s},P_{ps}$}]
    \item[$\hat{g}, \mathcal{\hat{G}}$] Index and set of strategic generators.
    \item[$g, \mathcal{G}$] Index and set of generators.
    \item[$r, \mathcal{R}$] Index and set of RES.
    \item[$s, \mathcal{S}$] Index and set of storage.
    \item[$t, \mathcal{T}, \textrm{t}^\textrm{end}$] Index, set of time horizon, and final time.
\end{IEEEdescription}

\vspace*{-4mm}
\subsection*{Parameters}
\begin{IEEEdescription}[\IEEEusemathlabelsep\IEEEsetlabelwidth{$P_{s},P_{ps}$}]
    \item[$\Delta^{\textrm{E}}_{\hat{g}} \hspace{-0.1mm}, \hspace{-0.5mm}  \Delta^{\textrm{H}}_{\hat{g}} \hspace{-0.1mm},\hspace{-0.5mm}  \Delta^{\textrm{PFR}}_{\hat{g}}$] \; \quad Length of interval for binary expansion.
    \item[$\Delta \textrm{f}_{\textrm{max}}$] Maximum admissible frequency deviation at nadir (Hz).
    \item[$\textrm{CF}_{rt}$] Capacity factor at hour $t$ for RES $r$.
    \vspace{0.5mm}
    \item[$\textrm{E}_{s}^\textrm{max} \hspace{-1mm}, \textrm{E}_{s}^\textrm{min}$]  Max/min state of charge for storage $s$ (MWh).
    \vspace{0.5mm}
    \item[$\textrm{Eff}_{s}^\textrm{c}, \textrm{Eff}_{s}^\textrm{d}$]   Charge and discharge efficiency for storage s.
    \item[$\textrm{f}_0$] Nominal power grid frequency (Hz).
    \vspace{0.5mm}
    \item[$\overline{\textrm{H}}_{\hat{g}}, \hspace{-0.5mm} \overline{\textrm{H}}_g, \hspace{-0.5mm} \overline{\textrm{H}}_s$]  Inertia constant of strategic generator $\hat{g}$, generator $g$, and storage $s$ (s).
    \vspace{0.5mm}
    \item[$\overline{\textrm{k}}^\textrm{E}_{\hat{g}}, \overline{\textrm{k}}^\textrm{H}_{\hat{g}}, \overline{\textrm{k}}^\textrm{PFR}_{\hat{g}}$] \; Multiplier limits for strategic generator $\hat{g}$. 
    \vspace{0.5mm}
    \item[$\textrm{L}^{\textrm{E}}_{\hat{g}} \hspace{-0.1mm}, \hspace{-0.5mm}  \textrm{L}^{\textrm{H}}_{\hat{g}} \hspace{-0.1mm},\hspace{-0.5mm}  \textrm{L}^{\textrm{PFR}}_{\hat{g}}$] \; Number of intervals for binary expansion.
    \vspace{0.5mm}
    \item[$\textrm{M}^{\lambda^\textrm{E}}_{\hat{g}} \hspace{-1.5mm},\hspace{-0.5mm}  \textrm{M}^{\lambda^\textrm{H}}_{\hat{g}} \hspace{-1.5mm},\hspace{-0.5mm}  \textrm{M}^{\lambda^\textrm{PFR}}_{\hat{g}}$] \qquad Large constants used for binary expansion of bilinear terms involving prices.
    \vspace{0.5mm}
    \item[$\textrm{M}^{\textrm{k}^\textrm{E}}_{\hat{g}} \hspace{-1mm},\hspace{-0.5mm}  \textrm{M}^{\textrm{k}^\textrm{H}}_{\hat{g}} \hspace{-1mm},\hspace{-0.5mm}  \textrm{M}^{\textrm{k}^\textrm{PFR}}_{\hat{g}}$] \qquad Large constants used for binary expansion of bilinear terms involving multipliers.
    \item[$\textrm{O}^\textrm{E}_{\hat{g}}, \textrm{O}^\textrm{E}_g, \textrm{O}^\textrm{E}_r, \textrm{O}^\textrm{E}_s$]  \qquad \; Energy price offer of strategic generator $\hat{g}$, generator $g$, RES $r$, and storage $s$ (\pounds/MWh). 
    \item[$\textrm{O}^\textrm{H}_{\hat{g}}, \textrm{O}^\textrm{H}_g, \textrm{O}^\textrm{H}_s$] \quad Inertia price offer of strategic generator $\hat{g}$, generator $g$, and storage $s$ (\pounds/MWs).
     \item[$\textrm{O}^\textrm{PFR}_{\hat{g}} \hspace{-1.5mm}, \textrm{O}^\textrm{PFR}_g \hspace{-1.5mm}, \textrm{O}^\textrm{PFR}_s \hspace{-1.5mm}, \textrm{O}^\textrm{EFR}_s$]  \qquad \qquad \; FR price offer of strategic generator $\hat{g}$, generator $g$, and storage $s$ (\pounds/MW).
    \vspace{0.5mm}
    \item[$\textrm{P}^\textrm{D}_t$] Total system demand at hour $t$ (MW).
    \item[$\textrm{PFR}^\textrm{max}_{\hat{g}} \hspace{-1.5mm}, \textrm{PFR}^\textrm{max}_g \hspace{-1.5mm}, \textrm{PFR}^\textrm{max}_s \hspace{-1.5mm}, \textrm{EFR}^\textrm{max}_s$] \quad\qquad\qquad\qquad\quad\; FR capacity of strategic generator $\hat{g}$, generator $g$ and storage $s$ (MW).
    \item[$\textrm{P}^\textrm{max}_{\hat{g}} \hspace{-1mm}, \hspace{-0.5mm}\textrm{P}^\textrm{max}_g \hspace{-1mm}, \hspace{-0.5mm}\textrm{P}^\textrm{max}_r \hspace{-1mm}, \hspace{-0.5mm}\textrm{P}^\textrm{max}_s$]  \qquad \;\;\; Rated power of strategic generator $\hat{g}$, generator $g$, RES $r$, and storage $s$ (MW).
    \item[$\textrm{P}^\textrm{msg}_{\hat{g}} \hspace{-1mm}, \hspace{-0.4mm}\textrm{P}^\textrm{msg}_{g}\hspace{-1mm}, \hspace{-0.4mm} \textrm{P}_{s}^\textrm{msg}$] \qquad  Minimum stable generation of strategic generator $\hat{g}$, generator $g$  and storage $s$ (MW).
    \item[$\textrm{RoCoF}_\textrm{max}$]   Maximum admissible RoCoF (Hz/s).
    \item[$\textrm{T}_\textrm{EFR},\hspace{-1mm} \textrm{T}_\textrm{PFR}$]   Delivery time of EFR and PFR (s).
    \vspace{0.5mm}
    \item[$\textrm{T}_g^\textrm{mdt}, \textrm{T}_g^\textrm{mut}, \textrm{T}_g^\textrm{st}$] \qquad Minimum down, minimum up, and start-up time for generator $g$ (h).
    \item[$\textrm{W}$] Penalty constant.
\end{IEEEdescription}

\vspace*{-5mm}
\subsection*{Primal variables \normalfont{(continuous and time dependent unless otherwise indicated)}} 
\begin{IEEEdescription}[\IEEEusemathlabelsep\IEEEsetlabelwidth{$P_{s},P_{ps}$}]
    \item[$b_{\hat{g}tn}^{\textrm{E}},  b_{\hat{g}tn}^{\textrm{H}},  b_{\hat{g}tn}^{\textrm{PFR}}$] \qquad Binary variables. Auxiliary variables used for binary expansion.
    \item[$E_{st}$] State of charge of storage $s$ (MWh).
    \vspace{0.5mm}
    \item[$H_{\hat{g}t}, \hspace{-0.5mm} H_{gt}, \hspace{-0.5mm} H_{st}$] \;\quad  Inertia from  strategic generator $\hat{g}$, generator $g$ and storage $s$ (MW$\cdot$s).
    \item[$H_{t}$]  Aggregate synchronous inertia (MW$\cdot$s). 
    \vspace{0.5mm}
    \item[$k^\textrm{E}_{\hat{g}t} \hspace{-0.1mm},\hspace{-0.5mm} k^\textrm{H}_{\hat{g}t} \hspace{-0.1mm},\hspace{-0.5mm} k^\textrm{PFR}_{\hat{g}t}$] \quad Bidding multipliers for price offers of strategic generator $\hat{g}$.
    \item[$P_{\hat{g}t}, \hspace{-0.5mm} P_{gt}, \hspace{-0.5mm} P_{rt}$] \quad Power produced by strategic generator $\hat{g}$, generator $g$, and RES $r$ (MW).
    \item[$P_{st}^\textrm{c}, \hspace{-0.5mm} P_{st}^\textrm{d}$]  Charge and discharge power of storage $s$ (MW).
    \vspace{0.5mm}
    \item[$P^\textrm{Loss}_{t}$] Largest power infeed (MW).
    \item[$PFR_{\hat{g}t},\hspace{-0.5mm} PFR_{gt}, \hspace{-0.5mm} PFR_{st},\hspace{-0.5mm} EFR_{st}$] \qquad\qquad\qquad\qquad\quad \;  FR from strategic generator $\hat{g}$, generator $g$, and storage $s$ (MW).
    \item[$PFR_{t}, EFR_{t}$]  \qquad  Aggregated provision of FR (MW).
    \vspace{0.5mm}
    \item[$y_{\hat{g}t}, y_{\hat{g}t}^\textrm{sd}, y_{\hat{g}t}^\textrm{sg}, y_{\hat{g}t}^\textrm{st},y_{gt}, y_{gt}^\textrm{sd}, y_{gt}^\textrm{sg}, y_{gt}^\textrm{st}$] \qquad\qquad\qquad\qquad\quad \; Binary variables. Commitment, shut-down, start generating, and start-up state of strategic generator $\hat{g}$ and generator $g$.
    \item[$y_{st}^\textrm{c}, y_{st}^\textrm{d}$]  Binary variables. Charging and discharging state of storage $s$.
    \vspace{1mm}
    \item[$z_{\hat{g}tn}^{\lambda^\textrm{E}} \hspace{-0.1mm},\hspace{-0.3mm}  z_{\hat{g}tn}^{\lambda^\textrm{H}} \hspace{-0.1mm},\hspace{-0.3mm}  z_{\hat{g}tn}^{\lambda^\textrm{PFR}}$]  \qquad Auxiliary variables used for binary expansion of bilinear terms involving prices.
    \vspace{1mm}
    \item[$z_{\hat{g}tn}^{{\textrm{k}^\textrm{E}}} \hspace{-0.1mm},\hspace{-0.3mm} z_{\hat{g}tn}^{{\textrm{k}^\textrm{H}}} \hspace{-0.1mm},\hspace{-0.3mm} z_{\hat{g}tn}^{{\textrm{k}^\textrm{PFR}}}$]  \qquad Auxiliary variables used for binary expansion of bilinear terms involving multipliers.
\end{IEEEdescription}

\vspace*{-5mm}
\subsection*{Dual variables \normalfont{(continuous and time dependent, where \textrm{(·)} represents a generic attribute)}}  
\begin{IEEEdescription}[\IEEEusemathlabelsep\IEEEsetlabelwidth{$P_{s},P_{ps}$}]
    \item[$\lambda^\textrm{E}_{t}$]  Energy price (£/MWh).
    \vspace{0.5mm}
    \item[$\lambda^\textrm{PFR}_{t}\hspace{-1mm}, \lambda^\textrm{EFR}_{t}$]  PFR and EFR prices (£/MW).
    \item[$\lambda_{t}^{\textrm{H}}$] Inertia price (£/MW$\textrm{s}$).
    \vspace{0.5mm}
    \item[$\mu_{\hat{g}t}^\textrm{(·)}, \hspace{-0.5mm} \mu_{gt}^\textrm{(·)}, \hspace{-0.5mm} \mu_{st}^\textrm{(·)}$] \quad Associated with commitment constraints of strategic generator $\hat{g}$, generator $g$, and charge and discharge state of storage $s$.
    \item[$\nu_{\hat{g}t}^\textrm{(·)}, \hspace{-0.5mm} \nu_{gt}^\textrm{(·)}, \hspace{-0.5mm} {\nu}^{\textrm{(·)}}_{rt}, \hspace{-0.5mm} \nu_{st}^\textrm{(·)}$] \qquad Associated with power constraints of strategic generator $\hat{g}$, generator $g$, RES $r$, and charge and discharge of storage $s$.
    \item[$\xi_{\hat{g}t}^\textrm{(·)}, \hspace{-0.5mm} \xi_{gt}^\textrm{(·)}, \hspace{-0.5mm} \xi_{st}^\textrm{(·)}$]  Associated with PFR constraints of strategic generator $\hat{g}$, generator $g$, and storage $s$.
    \item[$\rho_{st}^\textrm{(·)}$] Associated with EFR constraints  of storage $s$.
    \item[$\tau_{\hat{g}t}, \hspace{-1mm} \tau_{gt}, \hspace{-1mm} \tau_{st}$] Associated with inertia constraint of strategic generator $\hat{g}$, generator $g$, and storage $s$.
    \item[$\upsilon_{st}^\textrm{(·)}$] Associated with state of charge constraints of storage $s$.
    \item[$\omega_{t}^\textrm{(·)}$] Associated with frequency-containment constraints.
\end{IEEEdescription}

\vspace*{-10pt}
\section{Introduction}

\IEEEPARstart{C}{ompetition} is a crucial aspect that regulators monitor since the widespread liberalisation of the electricity sector in Europe, the Americas and Australasia. Whereas transmission and distribution remain regulated, generation and retail represent competitive markets \cite{green1992competition,Pollitt2024}. Notably, over the past decade, the generation sector has seen a notable increase in the number of players, including a relevant presence of Renewable Energy Sources (RES). This trend has reduced market concentration, improving competition levels \cite{beis2021competition,Serra2022}.

However, the vast integration of non-synchronous RES introduces several challenges for the secure operation of power systems. Since RES do not provide inertia, power systems have become more vulnerable to frequency deviations \cite{mohandes2019review}. 

Low inertia levels present significant operational challenges in electrical islands such as Great Britain (GB), Australia, and Texas. These power systems are not connected through AC links to any other system, so they must internally support sudden imbalances \cite{LuisCovid, MancarellaFragile}. The vulnerability of frequency stability has led to a diversified market for frequency-containment Ancillary Services (AS) on these systems, particularly for Frequency Response (FR) \cite{nside2023,Ela2019, AEMO2023a, AEMO2023b}. Meanwhile, most of the inertia provision relies on the rotating masses of synchronous generators, mostly provided as an energy by-product \cite{FRCR_NGESO_2023, AEMO2020power}. Discussions about the market design for inertia are still ongoing, and the main mechanism has relied on long-term competitive tenders \cite{appleby2019addressing, aemo_synchronous_2021, NG_Pathfinder2022}.

Synchronous generators remain indispensable in providing essential AS, such as Primary Frequency Response (PFR) and inertia. However, AS provision from synchronous generators varies significantly by turbine type. Steam turbine-based technologies, such as nuclear, inherently provide inertia due to their large rotating masses; however, their slow dynamics generally preclude them from offering PFR \cite{KundurBook}. In contrast, gas-fired generation, particularly those employing Combined-Cycle Gas Turbines (CCGTs) and Open-Cycle Gas Turbines (OCGTs), leverages governor control mechanisms, allowing them to be more flexible for providing frequency-containment AS \cite{OPFChavez, KundurBook}. 

Gas turbine-based generation serves as a `dispatchable' technology within the wholesale market as well as active participation in day-ahead and intra-day markets, which has a crucial role in providing backup energy supply for RES in the transition towards decarbonised electricity systems \cite{McKinsey2023}. Additionally, their unique ability to provide both inertia and PFR makes them a strong player in the AS market. 

Thus, in the current transition to decarbonisation, it is relevant to understand if there is room for these flexible units to inflict market power in a future market design that includes the energy market and a variety of services in the frequency-containment AS market, such as inertia and FR.

To effectively monitor and address market power in the electricity market, literature has focused on modelling market equilibria in the presence of \textit{strategic behaviour}. If certain players are aware that they can influence prices, the market will exhibit imperfect competition. Bi-level optimisation is the field of game theory that models a two-stage hierarchical games in which a \textit{leader} makes decisions before the \textit{followers} \cite{luo1996mathematical, dempe2003annotated}. In liberalised electricity market operation, the strategic behaviour can be analysed through the \textit{strategic bidding} choices of individual market players \cite{pozo2017basic}.

Strategic bidding has been extensively studied in electricity markets, where the main modelling tool is the Mathematical Program with Equilibrium Constraints (MPEC). To represent the bi-level problem, MPECs integrate the Lower Level (LL) problem, representing the competitive followers, into the Upper Level (UL) problem, representing the strategic leader game. The LL problem is transformed into its Karush–Kuhn–Tucker (KKT) conditions, which are added as constraints to the UL problem to get the final single level formulation. 

Relevant strategic bidding works can be found in \cite{Hobbs_2000, Pereira_2005, Bakirtzis2007, Ruiz2009}. An initial formulation was stated in \cite{Hobbs_2000}, which analyses the short-run price implications of market power. A more sophisticated formulation was done in \cite{Pereira_2005}, which considers uncertainties and a combination of quantities and prices for the strategic player. Two similar formulations were done in \cite{Bakirtzis2007, Ruiz2009}. In \cite{Bakirtzis2007}, step-wise offers consisting of multiple price-quantity pairs were developed, whereas \cite{Ruiz2009} extended it with locational marginal prices.

Given that KKT conditions only guarantee global optimality for convex LL problems, analysis of strategic bidding based on MPEC formulations usually derives simplified modelling of the LL problem \cite{pozo2017basic}. However, modelling non-convexities becomes critical when considering AS such as inertia and PFR, which strictly depend on the commitment status of synchronous generators, represented as binary variables in Unit Commitment (UC) problems, such as the Mixed-Integer Second-Order Cone Program (MISOCP) defined in \cite{LuisMultiFR, LuisSyntPricing, matamala2024cost}. 

An existing approach to solving bi-level problems with non-convexities is to use a primal-dual formulation. The primal-dual formulation allows the handling of primal and dual variables in the same optimisation problem. One of the first primal-dual formulations in electricity markets was presented in \cite{ruiz2012_primalDual} and was developed to price non-convexities while imposing nonnegative profits for market players. Its stochastic formulation was developed in \cite{Abbaspourtorbati_Conejo_2016}. A similar formulation that allows the provision of AS, such as regulation and reserve, was developed in \cite{goudarzi2021clearing}. 

Works that focus on pricing and directly apply a primal-dual formulation to solve bi-level problems include \cite{goudarzi2021energy, goudarzi2024strengthened, QIU2024122929}. These works assume a convex LL problem to simplify modelling, without addressing strategic bidding

In \cite{goudarzi2021energy} a pricing mechanism that ensures nonnegative profits is developed. The UL determines the commitment of generation and flexible demand, whereas the LL solves the economic dispatch. The representation as a single level problem is straightforward, considering that non-convex variables are defined in the UL. At the same time, the convex LL enables strong duality, thus simplifying dealing with bilinear terms in the nonnegative profit constraints. An improved version of \cite{goudarzi2021energy} is presented in \cite{goudarzi2024strengthened}, which considers FR provision to support frequency imbalances. This model uses Bender Cuts and Lagrangian Dual Decomposition to define a Strengthened Primal-Dual Decomposition that avoids the use of disjunctive parameters, used to linearise bilinear terms that appear in the nonnegative profit constraints, making the bi-level pricing problem solvable for a large number of scenarios. Authors in \cite{QIU2024122929} developed a primal-dual bi-level pricing model that considers a frequency-containment UC in the UL and a profit maximisation dispatch for virtual power plants in the LL. This work performs a primal-dual formulation for the UL while the convex LL is replaced by its KKT conditions.

In \cite{Ye_Papadaskalopoulos_Kazempour}, a methodology that models strategic bidding while considering non-convexities is presented. Through a strategic primal-dual bi-level formulation, \cite{Ye_Papadaskalopoulos_Kazempour} defines the UL problem as a profit maximisation for a strategic player. In contrast, the non-convex LL problem minimises system costs, represented by bidding offers from all market players. To derive a single level formulation, the non-convex LL is formulated as a primal-dual Duality Gap (DG) minimisation problem. Finally, the single level model maximises the strategic player profits while minimising the DG, which is penalised by a positive constant.

The analysis of market power in the coupled energy and frequency-containment AS market has not been appropriately addressed.  This gap stems from the simultaneous need to address: (i) a non-convex LL problem, key for dealing with AS such as inertia and PFR, that can not be expressed in terms of its KKT conditions; (ii) the complexity of dealing with non-linear frequency-containment constraints; and (iii) the challenge of handling bilinear terms that appear in the UL and LL problems.

In this work, a strategic market player's decision-making is modelled through a bi-level optimisation problem, following the methodology used in \cite{Ye_Papadaskalopoulos_Kazempour}, while considering the MISOCP formulation defined in \cite{LuisMultiFR, LuisSyntPricing, matamala2024cost}. The UL problem represents the profit maximisation of the strategic generator considering its participation in the energy and frequency-containment AS market. The LL problem represents the market-clearing process considering system-wide constraints such as meeting energy demand and AS provision. It is worth clarifying that this work focuses on under-frequency containment AS, given that frequency drops are the main concern in most grids. However, over-frequency events can be modelled symmetrically.

Specifically, the novel contributions of this work are:

\begin{enumerate}
    \item Develop a strategic primal-dual bi-level model based on a MISOCP formulation, enabling the computation of time-variant strategic decisions. 
    \item Present theoretical and quantitative evidence demonstrating the impact of strategic behaviour in the energy and AS market.
    \item Analyse a case study for the GB power system in 2030 that evidences the risks associated with market power.
\end{enumerate}

The remainder of this paper is structured as follows: Section~\ref{sec:Methodology} describes the mathematical methodology that models the strategic behaviour in the energy and AS market. Section~\ref{sec:Results} illustrates the model's applicability through a relevant case study representative of the GB power system in 2030. Finally, Section~\ref{sec:Conclusion} concludes.

\section{Modelling Approach} \label{sec:Methodology}

In this section, we develop a strategic primal-dual bi-level model inspired by the formulation in \cite{Ye_Papadaskalopoulos_Kazempour}. The model considers an electricity market represented as a multi-period frequency-secured UC with hourly resolution, formulated as an MISOCP optimisation model, based on the work done in \cite{LuisMultiFR, LuisSyntPricing, matamala2024cost}.

Regarding the mathematical notation, Latin letters are used for parameters, Latin \textit{italic} letters represent primal variables and Greek symbols in lowercase represent dual variables. Variables in parenthesis after semicolon represent dual variables associated with the corresponding constraint. We use the term \textit{system-wide constraints} for constraints related to the system requirements, such as frequency limits or balancing load and generation. In contrast, \textit{private constraints} are associated with specific market players' constraints. 

\subsection{Upper Level: Strategic Player Profits Maximisation} \label{sec:upperLevel}
The UL problem (\ref{eq:UpperLevel}) states the strategic player profits maximisation. Time-variant strategic decisions variables $k^\textrm{E}_{\hat{g}t}, k^\textrm{H}_{\hat{g}t}, k^\textrm{PFR}_{\hat{g}t}$ represent strategic multipliers of the bidding offers in the energy, inertia, and PFR markets, respectively.
\begin{subequations}\label{eq:UpperLevel}
\begin{alignat}{2}
    &  \displaystyle \max_{V_\textrm{UL}}  \Biggl\{  \sum_{t \in \mathcal{T}}  \Biggr[
     \sum_{\hat{g} \in \mathcal{\hat{G}}} \biggl(\lambda^\textrm{E}_{t}  P_{\hat{g}t}  + \lambda^\textrm{H}_{t}  H_{\hat{g}t}  + \lambda^\textrm{PFR}_{t}  PFR_{\hat{g}t}   &&\nonumber \\[-5pt]
    & \quad\qquad\qquad  - \textrm{O}^\textrm{E}_{\hat{g}} P_{\hat{g}t}  -  \textrm{O}^\textrm{H}_{\hat{g}}  H_{\hat{g}}  - \textrm{O}^\textrm{PFR}_{\hat{g}}  PFR_{\hat{g}t}  \biggl)  \Biggr]  \Biggl\} &&
    \label{eq:objFunc_strat}
\end{alignat}
\vspace*{-10mm}

\begin{alignat}{3}
    & 1 \leq k^\textrm{E}_{\hat{g}t} \leq \overline{\textrm{k}}^\textrm{E}_{\hat{g}} && \quad  \forall \hat{g}t  \label{eq:kE_strat} \\
    & 1 \leq k^\textrm{H}_{\hat{g}t} \leq \overline{\textrm{k}}^\textrm{H}_{\hat{g}} && \quad  \forall \hat{g}t  \label{eq:kH_strat} \\
    & 1 \leq k^\textrm{PFR}_{\hat{g}t} \leq \overline{\textrm{k}}^\textrm{PFR}_{\hat{g}} && \quad  \forall \hat{g}t  \label{eq:kPFR_strat} 
\end{alignat}
\vspace*{-10mm}

\begin{alignat}{3}
    & V_\textrm{UL}  = &&  \Bigl\{ k^\textrm{E}_{\hat{g}t} , k^\textrm{H}_{\hat{g}t}, k^\textrm{PFR}_{\hat{g}t}  \Bigl\}     \label{eq:strategic_vars} 
\end{alignat}
\end{subequations}

\subsection{Primal Lower Level: Frequency-Containment UC} \label{sec:FreqSecUC}
The primal LL problem (\ref{eq:LowerLevel}) considers the MISOCP formulation defined in \cite{LuisMultiFR, LuisSyntPricing, matamala2024cost}. The objective function (\ref{eq:objFunc}) minimises system costs associated with bidding offers.

\begin{subequations}\label{eq:LowerLevel}
\begin{alignat}{2}
    &  \displaystyle \min_{V_\textrm{LL}} \hspace{-0.5mm} \Biggl\{ \hspace{-0.3mm}  \sum_{t \in \mathcal{T}} \hspace{-0.5mm} \Biggr[ \hspace{-0.5mm}
     \sum_{\hat{g} \in \mathcal{\hat{G}}} \hspace{-0.5mm} \bigl( \hspace{-0.5mm}   \textrm{O}^\textrm{E}_{\hat{g}}  k^\textrm{E}_{\hat{g}t}  P_{\hat{g}t}  \hspace{-1mm} + \hspace{-1mm} \textrm{O}^\textrm{H}_{\hat{g}}  k^\textrm{H}_{\hat{g}t}  H_{\hat{g}t} \hspace{-1mm} + \hspace{-1mm} \textrm{O}^\textrm{PFR}_{\hat{g}}  k^\textrm{PFR}_{\hat{g}t} PFR_{\hat{g}t} \hspace{-0.5mm} \bigl) &&\nonumber \\[3pt]
    & + \hspace{-1.5mm} \sum_{g \in \mathcal{G}} \hspace{-0.5mm} \bigl( \hspace{-0.5mm}\textrm{O}^\textrm{E}_g P_{gt} \hspace{-0.5mm} + \hspace{-0.5mm} \textrm{O}^\textrm{H}_g  H_{gt} \hspace{-0.5mm} + \hspace{-0.5mm}\textrm{O}^\textrm{PFR}_g  PFR_{gt} \hspace{-0.5mm} \bigl) \; + \sum_{r \in \mathcal{R}}  \textrm{O}^\textrm{E}_r P_{rt}  &&\nonumber \\[-4pt]
    & +  \hspace{-1.5mm} \sum_{s \in \mathcal{S}} \hspace{-0.5mm} \bigl(\hspace{-0.5mm}  \textrm{O}^\textrm{E}_s P_{st}^\textrm{d}  \hspace{-1mm} + \hspace{-1mm} \textrm{O}^\textrm{H}_s  H_{st} \hspace{-1mm} + \hspace{-1mm} \textrm{O}^\textrm{PFR}_s  PFR_{st} \hspace{-1mm} + \hspace{-1mm} \textrm{O}^\textrm{EFR}_s EFR_{st} \hspace{-0.5mm} \bigl) \hspace{-0.8mm} \Biggr] \hspace{-0.5mm} \Biggl\} \hspace{-1.5mm} && 
    \label{eq:objFunc}
\end{alignat}

\vspace{-10pt}
\begin{alignat}{3}
    & V_\textrm{LL}  = && \Bigl\{ y_{\hat{g}t},y_{\hat{g}t}^\textrm{sg}, y_{\hat{g}t}^\textrm{sd},y_{\hat{g}t}^\textrm{st}, P_{\hat{g}t}, H_{\hat{g}t},PFR_{\hat{g}t}, \nonumber\\[3pt]  
    & && y_{gt},y_{gt}^\textrm{sg},y_{gt}^\textrm{sd},y_{gt}^\textrm{st}, P_{gt},H_{gt},PFR_{gt},   \nonumber\\[3pt] 
    & && P_{rt}, y_{st}^\textrm{c}, y_{st}^\textrm{d}, E_{st},  P_{st}^\textrm{c}, P_{st}^\textrm{d}, H_{st}, PFR_{st}, \nonumber\\[3pt] 
    & &&   EFR_{st}, P^\textrm{Loss}_{t}, H_{t}, PFR_{t}, EFR_{t}  \Bigl\}     \label{eq:primal_vars} 
\end{alignat}

\subsubsection{System-wide constraints} 
Power balance is defined in constraint (\ref{eq:e_balance}). Total synchronous inertia is defined by constraint (\ref{eq:SyncInertia}). PFR, provided by synchronous generators and Pump Hydro Energy Storage (PHES), is defined by constraint~(\ref{eq:pfr_total}), while EFR, provided by Battery Energy Storage Systems (BESS), is defined in (\ref{eq:efr_total}).

\vspace{-10pt}
\begin{alignat}{4}
    & \textrm{P}^\textrm{D}_t &&=  \sum_{\hat{g} \in \mathcal{\hat{G}}}P_{\hat{g}t} + \sum_{g \in \mathcal{G}}P_{gt} && \nonumber \\
    & &&  \quad + \sum_{r \in \mathcal{R}} P_{rt}  + \sum_{s \in \mathcal{S}} ( P_{st}^\textrm{d} -  P_{st}^\textrm{c} )     &&:  (\lambda_{t}^{\textrm{E}}) && \quad \forall t  \label{eq:e_balance} \\
    & H_{t}  &&=  \sum_{ \hat{g} \in \mathcal{\hat{G}} } H_{\hat{g}t} + \sum_{ g \in \mathcal{G} } H_{gt} + \sum_{ s \in \mathcal{S} } H_{st} &&:  (\lambda_{t}^\textrm{H}) && \quad \forall t \label{eq:SyncInertia}\\
    &PFR_{t}  &&=  \sum_{ \hat{g} \in \mathcal{\hat{G}} } PFR_{\hat{g}t}  && \nonumber \\
    & &&  \quad + \sum_{ g \in \mathcal{G} } PFR_{gt} + \sum_{ s \in \mathcal{S} } PFR_{st} &&:  (\lambda_{t}^\textrm{PFR})  && \quad \forall {t} \label{eq:pfr_total}\\
    &EFR_{t}  &&=  \sum_{ s \in \mathcal{S} } EFR_{st} &&:  (\lambda_{t}^\textrm{EFR})  &&  \quad  \forall {t} \label{eq:efr_total} \\
    &\textrm{P}_t^\textrm{Loss}  && \leq  P^\textrm{Loss}_{t} &&: (\omega^\textrm{Loss}_{t}) && \quad \forall t \label{eq:maxPloss}\\
    &  H_{t} && \geq  \frac{ P^\textrm{Loss}_{t} \cdot \textrm{f}_0}{2\cdot\textrm{RoCoF}_\textrm{max}  }  &&:(\omega^\textrm{RoCoF}_{t}) && \quad \forall t \label{eq:Rocof}\\
    & P^\textrm{Loss}_{t}  && \leq  EFR_{t} + PFR_{t}    &&:(\omega^\textrm{q-s-s}_{t})  && \quad \forall t \label{eq:qss}
\end{alignat}

\vspace{-20pt}
\begin{multline} \label{eq:nadirSOC}
    \norm{
    \setlength{\arraycolsep}{2.2pt}
    \begin{bmatrix} 
        \frac{1}{\textrm{f}_0}  
        & \frac{-\textrm{T}_\textrm{EFR}}{4\Delta \textrm{f}_{\textrm{max}}}  
        &  \frac{-1}{\textrm{T}_\textrm{PFR}} 
        & 0 \\[5pt]
        0 &\frac{-1}{\sqrt{\Delta \textrm{f}_{\textrm{max}}}} & 0 &\frac{1}{\sqrt{\Delta \textrm{f}_{\textrm{max}}}}
    \end{bmatrix}
    \begin{bmatrix} 
        H_{t}   
        \\ EFR_{t} 
        \\ PFR_{t} 
        \\  P^\textrm{Loss}_{t} 
    \end{bmatrix} 
    } 
    \leq
    \\
    \setlength{\arraycolsep}{2.2pt}
      \begin{bmatrix} 
        \frac{1}{\textrm{f}_0}  
        &\frac{-\textrm{T}_\textrm{EFR}}{4\Delta \textrm{f}_{\textrm{max}}}  
        &\frac{1}{\textrm{T}_\textrm{PFR}} 
        &0 
    \end{bmatrix} \hspace{-1.5mm}
    \begin{bmatrix} 
        H_{t}    
        \\ EFR_{t} 
        \\ PFR_{t} 
        \\  P^\textrm{Loss}_{t} 
    \end{bmatrix}  \hspace{-1.5mm}: \hspace{-1mm}(\omega_{t}^\textrm{nadir1}\hspace{-1mm},\omega_{t}^\textrm{nadir2}\hspace{-1mm},\omega_{t}^\textrm{nadir3}) \; \forall t
\end{multline}

The set of system-wide constraints that determine AS requirements for frequency containment are shown in eqs. (\ref{eq:maxPloss})-(\ref{eq:nadirSOC}). The decision variable $P^\textrm{Loss}_{t}$ represents the largest possible contingency which is limited by the maximum possible generation loss at each hour, considering $\textrm{P}_t^\textrm{Loss}=1.8 \textrm{GW}$ for GB in 2030 (see Table \ref{tab:GenerationMix}), and represented in (\ref{eq:maxPloss}). The Rate-of-Change-of-Frequency (RoCoF) constraint \eqref{eq:Rocof} guarantees that no islanding-protection scheme will be triggered, as this could exacerbate the frequency excursion. Quasi-steady-state (q-s-s) constraint (\ref{eq:qss}) defines that FR must be at least equal to the size of the outage to stabilise frequency at a value typically lower than nominal. Finally, nadir constraint (\ref{eq:nadirSOC}) avoids the activation of under-frequency load shedding and is formulated as a standard Second-Order Cone (SOC).
\begin{alignat}{3}
    & y_{\hat{g}t} = y_{\hat{g}t-1} + y_{\hat{g}t}^\textrm{sg} - y_{\hat{g}t}^\textrm{sd} &&: (\mu^{\textrm{rel}}_{\hat{g}t}) && \forall \hat{g}t\label{eq:CommitmentTime_s}\\[3pt]
    & y_{\hat{g}t}^\textrm{sg} = y_{\hat{g}t-\textrm{T}_{\hat{g}}^\textrm{st}}^\textrm{st} &&: (\mu^{\textrm{st}}_{\hat{g}t}) &&\forall \hat{g}t \label{eq:StartUpIndicator_s}\\
    & y_{\hat{g}t}^\textrm{st} \leq 1-y_{\hat{g}t-1} - \hspace{-4mm} \sum_{j=t-\textrm{T}_{\hat{g}}^\textrm{mdt}}^t \hspace{-3mm} y_{\hat{g}j}^\textrm{sd} &&: (\mu^{\textrm{mdt}}_{\hat{g}t}) &&  \forall \hat{g}t\label{eq:minDownTime_s}\\ 
    & y_{\hat{g}t}^\textrm{sd} \leq y_{\hat{g}t-1} - \hspace{-4mm} \sum_{j=t-\textrm{T}_{\hat{g}}^\textrm{mut}}^{t} \hspace{-3mm} y_{\hat{g}j}^\textrm{sg} &&: (\mu^{\textrm{mut}}_{\hat{g}t}) &&  \forall \hat{g}t\label{eq:minupTime_s}\\
    & y_{\hat{g}t}  \textrm{P}_{\hat{g}}^\textrm{msg}  \leq P_{\hat{g}t} \leq y_{\hat{g}t}  \textrm{P}_{\hat{g}}^\textrm{max} &&:  ({\nu}^{\textrm{min}}_{\hat{g}t}, {\nu}^{\textrm{max}}_{\hat{g}t})  &&  \forall \hat{g}t  \label{eq:gen_s}\\[3pt]
    & H_{\hat{g}t} =  \overline{\textrm{H}}_{\hat{g}} \; \textrm{P}^\textrm{max}_{\hat{g}} \; y_{\hat{g}t}   &&: (\tau_{\hat{g}t})   &&  \forall \hat{g}t  \label{eq:inertia_gen_s}\\[3pt]
    & 0 \leq PFR_{\hat{g}t} \leq y_{\hat{g}t}  \textrm{PFR}_{\hat{g}}^\textrm{max}  &&: (\xi^{\textrm{min}}_{\hat{g}t}, \xi^{\textrm{max1}}_{\hat{g}t}) &&  \forall \hat{g}t  \label{eq:pfr1_gen_s}  \\[3pt]
    & PFR_{\hat{g}t} \leq y_{\hat{g}t}  \textrm{P}_{\hat{g}}^\textrm{max} - P_{\hat{g}t} &&: (\xi^{\textrm{max2}}_{\hat{g}t})  &&   \forall \hat{g}t   \label{eq:pfr2_gen_s}\\[3pt]
    & y_{\hat{g}t},\; y_{\hat{g}t}^\textrm{st},\; y_{\hat{g}t}^\textrm{sg},\; y_{\hat{g}t}^\textrm{sd} \in  \{0, 1\} && && \forall \hat{g}t \label{eq:Binary_y_s} \\[3pt]
    & 0 \leq P_{rt} \leq \textrm{CF}_{rt}  \textrm{P}_{r}^\textrm{max} &&:  ({\nu}^{\textrm{min}}_{rt}, {\nu}^{\textrm{max}}_{rt}) &&  \forall rt  \label{eq:res_gen} \\
%
    & E_{s1}  = \textrm{E}_{s}^\textrm{ini}  &&:  ( \upsilon^{\textrm{ini}}_{s1}) && \forall s  \label{eq:e_stor_ini}\\[3pt]
    & E_{st^\textrm{end}}  = \textrm{E}_{s}^\textrm{end}  &&:  ( \upsilon^{\textrm{end}}_{st^\textrm{end}} ) && \forall s \label{eq:e_stor_end} \\[3pt]
    & y_{st}^\textrm{c}  \textrm{P}_{s}^\textrm{msg}  \leq P_{st}^\textrm{c} \leq y_{st}^\textrm{c} \textrm{P}_{s}^\textrm{max}    &&:   (\nu^{\textrm{min-c}}_{st} \hspace{-1.5mm}, \nu^{\textrm{max-c}}_{st}) && \forall st \label{eq:genc} \\[3pt]
    & y_{st}^\textrm{d}  \textrm{P}_{s}^\textrm{msg}  \leq P_{st}^\textrm{d} \leq y_{st}^\textrm{d}  \textrm{P}_{s}^\textrm{max}  &&:   (\nu^{\textrm{min-d}}_{st} \hspace{-1.5mm} , \nu^{\textrm{max-d}}_{st}) && \forall st \label{eq:gend}\\[3pt]
    & H_{st} = \overline{\textrm{H}}_s \; \textrm{P}^\textrm{max}_s \; (y_{st}^\textrm{d} + y_{st}^\textrm{c})  &&:  (\tau_{st}) && \forall st \label{eq:H_stor} \\[3pt]
    & 0 \leq PFR_{st} \leq y_{st}^\textrm{d}  \textrm{PFR}_{s}^\textrm{max}   &&: (\xi^{\textrm{min}}_{st}, \xi^{\textrm{max1}}_{st}) && \forall st \label{eq:pfr1_stor} \\[3pt]
    & PFR_{st} \leq y_{st}^\textrm{d} \textrm{P}_{s}^\textrm{max} \hspace{-1mm} - \hspace{-1mm} P_{st}^\textrm{d} +  P_{st}^\textrm{c}  &&: (\xi^{\textrm{max2}}_{st}) && \forall st \label{eq:pfr2_stor} \\[3pt]
    & 0 \leq EFR_{st} \label{eq:efr0_stor}  &&: (\rho^{\textrm{min}}_{st}) && \forall st \\[3pt]
    & EFR_{st} \leq  (y_{st}^\textrm{c} + y_{st}^\textrm{d})  \textrm{EFR}_{s}^\textrm{max} \label{eq:efr1_stor}  &&: (\rho^{\textrm{max1}}_{st}) && \forall st \\[3pt]
    & EFR_{st}\leq ( y_{st}^\textrm{d}+y_{st}^\textrm{c})  \textrm{P}_{s}^\textrm{max}  && &&  \nonumber\\
    & \qquad \qquad- P_{st}^\textrm{d} + P_{st}^\textrm{c} &&: (\rho^{\textrm{max2}}_{st}) &&  \forall st \label{eq:efr2_stor}\\[3pt]
    & \textrm{E}_{s}^\textrm{min} \leq E_{st} \leq \textrm{E}_{s}^\textrm{max}  &&:  ( \upsilon^{\textrm{min}}_{st}, \upsilon^{\textrm{max}}_{st} ) && \forall st \label{eq:e_stor_minmax}\\[3pt]
    & E_{st} \hspace{-1mm} = \hspace{-1mm}E_{s{t-1}} \hspace{-1mm} + \hspace{-1mm}P_{st}^\textrm{c} \textrm{Eff}_{s}^\textrm{c}  \hspace{-1mm}  -  \hspace{-1mm} P_{st}^\textrm{d}/ \textrm{Eff}_{s}^\textrm{d}   &&:  (\upsilon^{\textrm{rel}}_{st})  && \forall st \label{eq:e_stor} \\[3pt]
    & y_{st}^\textrm{c} + y_{st}^\textrm{d} \leq 1  &&:  (\mu^{\textrm{rel}}_{st} ) && \forall st \label{eq:Sum_Binary_ys}\\[3pt]
    & y_{st}^\textrm{c},\; y_{st}^\textrm{d} \in  \{0, 1\} && && \forall st \label{eq:Binary_ys} 
\end{alignat}

\subsubsection{Private constraints }
Thermal generators are defined by the private constraints~(\ref{eq:CommitmentTime_s})-(\ref{eq:Binary_y_s}), which are defined $\forall \hat{g} \in \mathcal{\hat{G}}, t \in  \mathcal{T}$ in the case of the strategic player but also apply $\forall g \in \mathcal{G}, t \in  \mathcal{T}$ in the case of the competitive thermal players. Constraint~(\ref{eq:CommitmentTime_s}) establishes the relation between online, start-generating and shut-down units at each hour. Constraint~(\ref{eq:StartUpIndicator_s}) represents that a unit starts generating if it was started up $\textrm{T}_g^\textrm{st}$ hours before. Minimum down and up times are modelled by constraints (\ref{eq:minDownTime_s}) and (\ref{eq:minupTime_s}), respectively. Constraint~(\ref{eq:gen_s}) defines the generation limits. Constraint~(\ref{eq:inertia_gen_s}) represents the inertia provision. Constraints (\ref{eq:pfr1_gen_s}) and (\ref{eq:pfr2_gen_s}) represent the maximum PFR deliverable and the available margin to provide it, respectively. Finally, constraint (\ref{eq:Binary_y_s}) imposes the binary nature of decision variables associated with commitment, start-up, start-generating, and shut-down. 

RES generation are modelled with an hourly generation profile as shown in private constraint~(\ref{eq:res_gen}), which applies $\forall r \in \mathcal{R}, t \in  \mathcal{T}$. 

Energy storage units are defined by private constraints~(\ref{eq:e_stor_ini})-(\ref{eq:Binary_ys}). Constraints~(\ref{eq:e_stor_ini})-(\ref{eq:e_stor_end}) apply $\forall s \in \mathcal{S}$, while constraints~(\ref{eq:genc})-(\ref{eq:Binary_ys}) apply $\forall s \in \mathcal{S}, t \in  \mathcal{T}$.  Constraints (\ref{eq:e_stor_ini}) and (\ref{eq:e_stor_end}) enforce the initial and final state of charge, respectively. Charge and discharge limits are modelled with constraints (\ref{eq:genc}) and (\ref{eq:gend}), respectively. State of charge is modelled through (\ref{eq:e_stor_minmax})-(\ref{eq:e_stor}). Constraints~(\ref{eq:Sum_Binary_ys})-(\ref{eq:Binary_ys}) state that the storage unit can be either charging or discharging, but not in both states simultaneously. PHES units can provide inertia in any state, whether they are discharging or charging, as defined in (\ref{eq:H_stor}), while the maximum and available margin for PFR are defined by constraints (\ref{eq:pfr1_stor}) and (\ref{eq:pfr2_stor}), respectively. In the case of BESS, minimum, maximum and available EFR margins are determined by constraints (\ref{eq:efr0_stor}), (\ref{eq:efr1_stor}) and (\ref{eq:efr2_stor}), respectively.

\subsection{Dual Lower Level} \label{sec:DualFreqSecUC}
In this section,  we derive the dual of the LL problem defined in section \ref{sec:FreqSecUC}. To do this, the primal LL problem (\ref{eq:LowerLevel}) is convexified by relaxing binary variables (\ref{eq:Binary_y_s}) and (\ref{eq:Binary_ys}), which are re-stated in (\ref{eq:Continuous_y_min_s})-(\ref{eq:Continuous_y__maxStrat}) for the strategic and competitive generators, and (\ref{eq:Continuous_ys_min})-(\ref{eq:Continuous_ys_max}) for storage units.

\begin{alignat}{3}
    &  0 \hspace{-1mm} \leq \hspace{-1mm} y_{\hat{g}t}, y_{\hat{g}t}^\textrm{st} , y_{\hat{g}t}^\textrm{sg}, y_{\hat{g}t}^\textrm{sd} \hspace{-1mm} &&:\hspace{-0.5mm}  (\mu^{\textrm{min}}_{\hat{g}t}\hspace{-0.5mm},\hspace{-0.5mm}\mu^{\textrm{min-st}}_{\hat{g}t}\hspace{-2mm}, \mu^{\textrm{min-sg}}_{\hat{g}t}\hspace{-2mm}, \mu^{\textrm{min-sd}}_{\hat{g}t}  )      \hspace{-1mm}     &&  \forall \hat{g}t \label{eq:Continuous_y_min_s} \\[3pt]
    &  y_{\hat{g}t}, y_{\hat{g}t}^\textrm{st}, y_{\hat{g}t}^\textrm{sg} , y_{\hat{g}t}^\textrm{sd} \hspace{-1mm} \leq  \hspace{-1mm} 1 \hspace{-1mm} &&: \hspace{-0.5mm} (\mu^{\textrm{max}}_{\hat{g}t} \hspace{-0.5mm},\hspace{-0.5mm} \mu^{\textrm{max-st}}_{\hat{g}t}\hspace{-2mm}, \mu^{\textrm{max-sg}}_{\hat{g}t}\hspace{-2mm}, \mu^{\textrm{max-sd}}_{\hat{g}t})   &&  \forall \hat{g}t \label{eq:Continuous_y__maxStrat} \\[3pt]
    & 0 \leq y_{st}^\textrm{c}, y_{st}^\textrm{d}    &&:    (\mu^{\textrm{min-c}}_{st}, \mu^{\textrm{min-d}}_{st} )           &&  \forall st \label{eq:Continuous_ys_min} \\[3pt]
    & y_{st}^\textrm{c}, y_{st}^\textrm{d}  \leq 1    &&:  (\mu^{\textrm{max-c}}_{st}, \mu^{\textrm{max-d}}_{st} )           &&  \forall st \label{eq:Continuous_ys_max} 
\end{alignat}
\end{subequations}

The methodology defined in \cite{andersen2002notes} is used to create the dual model. After the relaxation, the primal MISOCP becomes a Second-Order Cone Program (SOCP), whose dual SOCP is stated in (\ref{eq:DualLowerLevel}). The dual LL objective function is stated in \eqref{eq:dual_objFunc}, while the list of variables is shown in \eqref{eq:dual_vars}. Constraints (\ref{eq:dual_y_k})-(\ref{dual_H_k}) represent dual constraints with respect to primal variables $y_{\hat{g}t}, y_{\hat{g}t}^\textrm{sg}, y_{\hat{g}t}^\textrm{sd}, y_{\hat{g}t}^\textrm{st},P_{\hat{g}t},PFR_{\hat{g}t},H_{\hat{g}t}$ for the strategic generator. Constraints (\ref{eq:dual_y_g})-(\ref{dual_H_g}) do the same for competitive generators. Constraint (\ref{dual_x_r}) represents the dual constraint with respect to RES generation $P_{rt}$. 
\begin{subequations}\label{eq:DualLowerLevel}
\begin{alignat}{2}
    &  \displaystyle \max_{V_\textrm{DLL}}  \Biggl\{ \sum_{t \in \mathcal{T}}  \Biggr[-
     \sum_{\hat{g} \in \mathcal{\hat{G}} } \bigl( \mu^{\textrm{max}}_{\hat{g}t} + \mu^{\textrm{max-st}}_{\hat{g}t} + \mu^{\textrm{max-sg}}_{\hat{g}t} + \mu^{\textrm{max-sd}}_{\hat{g}t} \bigl) &&\nonumber \\
    &  - \hspace{-1mm} \sum_{g \in \mathcal{G}} \hspace{-0.5mm} \bigl(  \mu^{\textrm{max}}_{gt} + \mu^{\textrm{max-st}}_{gt} + \mu^{\textrm{max-sg}}_{gt} + \mu^{\textrm{max-sd}}_{gt}     \bigl)   - \hspace{-1mm} \sum_{r \in \mathcal{R}} \hspace{-1mm}  {\nu}^{\textrm{max}}_{rt} \textrm{CF}_{rt} \textrm{P}_{r}^\textrm{max} &&\nonumber \\
    & - \hspace{-1mm} \sum_{s \in \mathcal{S}} \hspace{-0.5mm}  \bigl( \mu^{\textrm{max-c}}_{st} +  \mu^{\textrm{max-d}}_{st} + \mu^{\textrm{rel}}_{st}  -\upsilon^{\textrm{min}}_{st}  \textrm{E}_{s}^\textrm{min}  +  \upsilon^{\textrm{max}}_{st}  \textrm{E}_{s}^\textrm{max}   \bigl)  &&\nonumber \\[-5pt]
    & + \hspace{-1mm} \textrm{P}_{t}^\textrm{Loss}  \omega^\textrm{Loss}_{t} + \textrm{P}^\textrm{D}_t  \lambda_{t}^{\textrm{E}}    \Biggr]  \hspace{-1mm}
    - \hspace{-1mm} \sum_{s \in \mathcal{S}} \bigl(\textrm{E}_{s}^\textrm{ini}  \upsilon^{\textrm{ini}}_{s1}  +\textrm{E}_{s}^\textrm{end} \upsilon^{\textrm{end}}_{st^\textrm{end}}  \bigl) \hspace{-1mm}
    \Biggl\} &&
    \label{eq:dual_objFunc} 
\end{alignat}

\vspace*{-5mm}

\begin{alignat}{3}
    &  V_\textrm{DLL} = && \Bigl\{ \mu_{gt}^\textrm{(·)},\nu_{gt}^\textrm{(·)}, \xi_{gt}^\textrm{(·)}, \tau_{gt}, {\nu}^{\textrm{(·)}}_{rt},  \mu_{st}^\textrm{(·)},\nu_{st}^\textrm{(·)}, \xi_{st}^\textrm{(·)}, \nonumber\\[3pt] 
    & && \rho_{st}^\textrm{(·)}, \tau_{st}, \upsilon_{st}^\textrm{(·)}, \omega_{t}^\textrm{(·)},  \lambda_{t}^{\textrm{E}}, \lambda_{t}^\textrm{H}, \lambda_{t}^\textrm{PFR}, \lambda_{t}^\textrm{EFR}     \Bigl\}     \label{eq:dual_vars}     
\end{alignat}

\vspace*{-10mm}

\begin{alignat}{3}    
    & \mu^{\textrm{min}}_{\hat{g}t} - \mu^{\textrm{max}}_{\hat{g}t} - {\nu}^{\textrm{min}}_{\hat{g}t} \textrm{P}_{\hat{g}}^\textrm{msg} + {\nu}^{\textrm{max}}_{\hat{g}t} \textrm{P}_{\hat{g}}^\textrm{max}  + \mu^{\textrm{rel}}_{\hat{g}t}   && && \nonumber\\
    & -  \mu^{\textrm{rel}}_{\hat{g}t+1} - \mu^{\textrm{mdt}}_{\hat{g}t+1}  +  \mu^{\textrm{mut}}_{\hat{g}t+1} + \xi^{\textrm{max1}}_{\hat{g}t} \textrm{PFR}_{\hat{g}}^\textrm{max}   && && \nonumber\\
    & +  \xi^{\textrm{max2}}_{\hat{g}t} \textrm{P}_{\hat{g}}^\textrm{max}  +  \tau_{\hat{g}t} \overline{\textrm{H}}_{\hat{g}}  \textrm{P}^\textrm{max}_{\hat{g}}           = 0      &&  \forall \hat{g}t &&\label{eq:dual_y_k} \\ 
    & \mu^{\textrm{min-sg}}_{\hat{g}t} -   \mu^{\textrm{max-sg}}_{\hat{g}t}  -  \mu^{\textrm{rel}}_{\hat{g}t}  +  \mu^{\textrm{st}}_{\hat{g}t}  - \hspace{-3mm} \sum_{j = t}^{t+\textrm{T}_{\hat{g}}^\textrm{mut}}  \mu^{\textrm{mut}}_{\hat{g}j}   = 0   && \forall \hat{g}t  && \label{eq:dual_ysg_k_I2} \\ 
    & \mu^{\textrm{min-sd}}_{\hat{g}t} - \mu^{\textrm{max-sd}}_{\hat{g}t} + \mu^{\textrm{rel}}_{\hat{g}t} - \mu^{\textrm{mut}}_{\hat{g}t} - \hspace{-3mm} \sum_{j = t}^{t+\textrm{T}_{\hat{g}}^\textrm{mdt}} \mu^{\textrm{mdt}}_{\hat{g}j}   = 0  \; && \forall \hat{g}t && \label{dual_ysd_k_tmDT} \\[3pt]
    & \mu^{\textrm{min-st}}_{\hat{g}t} - \mu^{\textrm{max-st}}_{\hat{g}t} - \mu^{\textrm{st}}_{\hat{g}t+\textrm{T}_g^\textrm{st}} - \mu^{\textrm{mdt}}_{\hat{g}t}  =  0  \quad && \forall \hat{g}t && \label{dual_yst_k} \\[3pt]
    & {\nu}^{\textrm{min}}_{\hat{g}t} - {\nu}^{\textrm{max}}_{\hat{g}t} - \xi^{\textrm{max2}}_{\hat{g}t} +\lambda^\textrm{E}_{t} =  k^\textrm{E}_{\hat{g}t} \; \textrm{O}^\textrm{E}_{\hat{g}}  \;\; && \forall \hat{g}t && \label{dual_x_k} \\[3pt]  
    & \xi^{\textrm{min}}_{\hat{g}t} - \xi^{\textrm{max1}}_{\hat{g}t} - \xi^{\textrm{max2}}_{\hat{g}t} + \lambda_{t}^\textrm{PFR} = k^\textrm{PFR}_{\hat{g}t} \; \textrm{O}^\textrm{PFR}_{\hat{g}}  \;\; && \forall \hat{g}t && \label{dual_PFRl_k} \\[3pt] 
    & -\tau_{\hat{g}t} + \lambda_{t}^{\textrm{H}} = k^\textrm{H}_{\hat{g}t} \;  \textrm{O}^\textrm{H}_{\hat{g}}   && \forall \hat{g}t && \label{dual_H_k}  \\[3pt] 
%
    & \mu^{\textrm{min}}_{gt} - \mu^{\textrm{max}}_{gt} - {\nu}^{\textrm{min}}_{gt} \textrm{P}_{g}^\textrm{msg} + {\nu}^{\textrm{max}}_{gt} \textrm{P}_{g}^\textrm{max}  + \mu^{\textrm{rel}}_{gt}   && && \nonumber\\
    & -  \mu^{\textrm{rel}}_{gt+1} - \mu^{\textrm{mdt}}_{gt+1}  +  \mu^{\textrm{mut}}_{gt+1} + \xi^{\textrm{max1}}_{gt} \textrm{PFR}_{g}^\textrm{max}   && && \nonumber\\
    & +  \xi^{\textrm{max2}}_{gt} \textrm{P}_{g}^\textrm{max}  +  \tau_{gt} \overline{\textrm{H}}_{g}  \textrm{P}^\textrm{max}_{g} = 0      &&  \forall gt &&\label{eq:dual_y_g} \\ 
    & \mu^{\textrm{min-sg}}_{gt} -   \mu^{\textrm{max-sg}}_{gt}  -  \mu^{\textrm{rel}}_{gt}  +  \mu^{\textrm{st}}_{gt}  - \hspace{-3mm} \sum_{j = t}^{t+\textrm{T}_{g}^\textrm{mut}}  \mu^{\textrm{mut}}_{gj}   = 0   && \forall gt  && \label{eq:dual_ysg_I2_g} \\ 
    & \mu^{\textrm{min-sd}}_{gt} - \mu^{\textrm{max-sd}}_{gt} + \mu^{\textrm{rel}}_{gt} - \mu^{\textrm{mut}}_{gt} - \hspace{-3mm} \sum_{j = t}^{t+\textrm{T}_{g}^\textrm{mdt}} \mu^{\textrm{mdt}}_{gj}   = 0  \; && \forall gt && \label{dual_ysd_tmDT_g} \\[3pt]
    & \mu^{\textrm{min-st}}_{gt} - \mu^{\textrm{max-st}}_{gt} - \mu^{\textrm{st}}_{gt+\textrm{T}_g^\textrm{st}} - \mu^{\textrm{mdt}}_{gt}  =  0  \quad && \forall gt && \label{dual_yst_g} \\[3pt]
    & {\nu}^{\textrm{min}}_{gt} - {\nu}^{\textrm{max}}_{gt} - \xi^{\textrm{max2}}_{gt} +\lambda^\textrm{E}_{t} =  \textrm{O}^\textrm{E}_{g}  \;\; && \forall gt && \label{dual_x_g} \\[3pt] 
    & \xi^{\textrm{min}}_{gt} - \xi^{\textrm{max1}}_{gt} - \xi^{\textrm{max2}}_{gt} + \lambda_{t}^\textrm{PFR} =  \textrm{O}^\textrm{PFR}_{g}  \;\; && \forall gt && \label{dual_PFRl_g} \\[3pt]  
    & -\tau_{gt} + \lambda_{t}^{\textrm{H}} =  \textrm{O}^\textrm{H}_{g}   && \forall gt && \label{dual_H_g} \\[3pt]
    &  {\nu}^{\textrm{min}}_{rt} - {\nu}^{\textrm{max}}_{rt} + \lambda^\textrm{E}_{t} =  \textrm{O}^\textrm{E}_{r}  \qquad\qquad\qquad\qquad && \forall rt && \label{dual_x_r} \\[3pt] 
%
    & \mu^{\textrm{min-d}}_{st} \hspace{-1.5mm} - \hspace{-1mm} \mu^{\textrm{max-d}}_{st} \hspace{-1mm} - \hspace{-1mm} \mu^{\textrm{rel}}_{st} \hspace{-1mm} -  \nu^{\textrm{min-d}}_{st} \textrm{P}_{s}^\textrm{msg}  +  \xi^{\textrm{max1}}_{st} \textrm{PFR}^\textrm{max}_s  &&    &&  \nonumber\\
    & + \rho^{\textrm{max1}}_{st} \textrm{EFR}^\textrm{max}_s + (\nu^{\textrm{max-d}}_{st}+ \xi^{\textrm{max2}}_{st}  +   \rho^{\textrm{max2}}_{st} && &&  \nonumber\\
    & + \tau_{st} \overline{\textrm{H}}_s) \textrm{P}^\textrm{max}_s  =   0                                                                  &&\forall st && \label{dual_yd_s} \\[3pt]
    & \mu^{\textrm{min-c}}_{st} \hspace{-1.5mm} - \hspace{-1mm} \mu^{\textrm{max-c}}_{st} \hspace{-1mm} - \hspace{-1mm} \mu^{\textrm{rel}}_{st} \hspace{-1mm} -  \nu^{\textrm{min-c}}_{st} \textrm{P}_{s}^\textrm{msg}   +     \rho^{\textrm{max1}}_{st} \textrm{EFR}^\textrm{max}_s && &&  \nonumber\\
    & + (\nu^{\textrm{max-c}}_{st}  +   \rho^{\textrm{max2}}_{st} + \tau_{st} \overline{\textrm{H}}_s) \textrm{P}^\textrm{max}_s  =   0   && \forall st && \label{dual_yc_s} \\[3pt]
    & \upsilon^{\textrm{min}}_{st} - \upsilon^{\textrm{max}}_{st} - \upsilon^{\textrm{rel}}_{st} + \upsilon^{\textrm{rel}}_{st+1}  = 0    && \forall st && \label{dual_es} \\[3pt]
    & \nu^{\textrm{min-d}}_{st}   -   \nu^{\textrm{max-d}}_{st}   -   \upsilon^{\textrm{rel}}_{st} / \textrm{Eff}_{s}^\textrm{d}  -  \xi^{\textrm{max2}}_{st}  -  \rho^{\textrm{max2}}_{st}  &&  &&  \nonumber\\
    & +  \lambda_{t}^{\textrm{E}}   =  \textrm{O}^\textrm{E}_s     && \forall st && \label{dual_xd_s} \\[3pt] 
    & \nu^{\textrm{min-c}}_{st}   -   \nu^{\textrm{max-c}}_{st}   -   \upsilon^{\textrm{rel}}_{st} \textrm{Eff}_{s}^\textrm{c}  +  \rho^{\textrm{max2}}_{st}  +  \lambda_{t}^{\textrm{E}}   =  0  \;\;  && \forall st && \label{dual_xc_s} \\[3pt]
    & \xi^{\textrm{min}}_{st}  -  \xi^{\textrm{max1}}_{st}  -  \xi^{\textrm{max2}}_{st}  +  \lambda_{t}^\textrm{PFR}  = \textrm{O}^\textrm{PFR}_s       && \forall st && \label{dual_PFRl_s} \\[3pt] 
    & \rho^{\textrm{min}}_{st} - \rho^{\textrm{max1}}_{st} -  \rho^{\textrm{max2}}_{st} + \lambda_{t}^\textrm{EFR}  =  \textrm{O}^\textrm{EFR}_s      && \forall st && \label{dual_EFRl_s} \\[3pt]  
    & - \tau_{st} + \lambda_{t}^\textrm{H} = \textrm{O}^\textrm{H}_s      && \forall st && \label{dual_H_s} \\[3pt] 
%
    & \omega^\textrm{q-s-s}_{t} + \frac{-\omega_{t}^\textrm{nadir1} + \omega_{t}^\textrm{nadir3}}{\textrm{T}_\textrm{PFR}}  - \lambda_{t}^\textrm{PFR} = 0 \;\quad && \forall t && \label{dual_PFRl} \\[3pt]
    & \omega^\textrm{q-s-s}_{t} - \frac{(\omega_{t}^\textrm{nadir1} + \omega_{t}^\textrm{nadir3})\textrm{T}_\textrm{EFR}}{4\Delta \textrm{f}_{\textrm{max}}}  -  \frac{\omega_{t}^\textrm{nadir2}}{\sqrt{\Delta \textrm{f}_{\textrm{max}}}} && &&  \nonumber\\
    & - \lambda_{t}^\textrm{EFR} = 0  && \forall t && \label{dual_EFRl} 
\end{alignat}

\vspace*{-10mm}
\begin{alignat}{3}    
    & \omega^\textrm{RoCoF}_{t} + \frac{\omega_{t}^\textrm{nadir1} + \omega_{t}^\textrm{nadir3}}{\textrm{f}_0} - \lambda_{t}^\textrm{H} = 0  \; && \forall t && \label{dual_H} \\[3pt]
    & \omega^\textrm{Loss}_{t} - \omega^\textrm{q-s-s}_{t} - \frac{\omega^\textrm{RoCoF}_{t} \textrm{f}_0}{2 \textrm{RoCoF}_\textrm{max}}  + \frac{\omega_{t}^\textrm{nadir2}}{\sqrt{\Delta \textrm{f}_{\textrm{max}}}} = 0 \;\;\; && \forall t && \label{dual_Ploss} \\[3pt]
    &  \norm{\begin{bmatrix} \omega_{t}^\textrm{nadir1} \\ \omega_{t}^\textrm{nadir2} \end{bmatrix}} \leq   \omega_{t}^\textrm{nadir3}  \; && \forall t && \label{dual_nadir_low} 
\end{alignat}
\end{subequations}

Constraints (\ref{dual_yd_s})-(\ref{dual_H_s}) represent dual constraint regarding $y_{st}^\textrm{d}, y_{st}^\textrm{c}, E_{st}, P_{st}^\textrm{d}, P_{st}^\textrm{c}, PFR_{st}, EFR_{st}, H_{st}$, respectively. Constraints (\ref{dual_PFRl})-(\ref{dual_Ploss}) are the dual constraints of the AS market variables $PFR_{t}, EFR_{t}, H_{t}, P^\textrm{Loss}_{t}$. Finally, constraint (\ref{dual_nadir_low}) is included to comply with the dual feasibility of the SOCP. All dual variables associated with inequality constraints in the primal LL problem are positive, whereas those associated with equality constraints can assume any continuous value.

\subsection{Single Level Primal-Dual Bi-Level Formulation} \label{ssec:singleLevel}
To address the strategic bidding problem, this work converts the bi-level model into a single level optimisation problem using the Penalty Function approach presented in \cite{Ye_Papadaskalopoulos_Kazempour} and \cite{penalty_biLevel}. This methodology allows the modelling of strategic decisions defined in the UL problem (\ref{eq:UpperLevel}), together with the system operator's LL problem (\ref{eq:LowerLevel}), considering that for a given UL variable set $V_\textrm{UL}$, the LL problem will set its optimal values $V_\textrm{LL}$ when the DG is minimal. As the DG represents the difference between the non-convex primal LL (\ref{eq:objFunc}) and the DLL (\ref{eq:dual_objFunc}) objective functions, it can not be zero. However, its value has to be small to provide a realistic market-clearing solution. 

The coupled maximisation of the UL objective function (\ref{eq:objFunc_strat}) alongside the minimisation of the DG is represented in the Single Level Primal-Dual Bi-Level Formulation (\ref{eq:SingleLevel}). The model incorporates a penalty parameter W, which governs these objectives' trade-offs. A small value of W will prioritise the profit maximisation of the strategic player but will provide a less realistic market-clearing solution for the LL. Conversely, a larger W will provide a more accurate LL solution with reduced profitability for the strategic player.

\vspace*{-5mm}

\begin{subequations}\label{eq:SingleLevel}

\begin{alignat}{2}
    \max_{V} &\quad \biggl\{  \eqref{eq:objFunc_strat} - \textrm{W} \cdot DG \biggl\} \label{eq:obj_singleLevel} \\[3pt]
    \text{where:} &\quad DG = \eqref{eq:objFunc} - \eqref{eq:dual_objFunc} \label{eq:DG_singleLevel} \\[3pt]
    \text{and:} &\quad V = \left\{  V_\textrm{UL}, V_\textrm{LL}, V_\textrm{DLL} \right\} \label{eq:var_singleLevel} \\[3pt]
    \text{subject to:} &\quad \eqref{eq:kE_strat} - \eqref{eq:kPFR_strat} \textrm{,} \; \eqref{eq:e_balance} - \eqref{eq:Continuous_ys_max} \textrm{, and} \; \eqref{eq:dual_y_k} - \eqref{dual_nadir_low}  \label{eq:constr_singleLevel}
\end{alignat}

Problem \eqref{eq:SingleLevel} is non-linear because of the SOC constraints \eqref{eq:nadirSOC} and \eqref{dual_nadir_low}; also, it contains bilinear terms associated with multiplication of variables that are expressed in \eqref{eq:objFunc_strat} and \eqref{eq:objFunc}, such as $\lambda^\textrm{E}_{t}  P_{\hat{g}t} , \lambda^\textrm{H}_{t}  H_{\hat{g}t}, \lambda^\textrm{PFR}_{t}  PFR_{\hat{g}t}$ and $k^\textrm{E}_{\hat{g}t}  P_{\hat{g}t},  k^\textrm{H}_{\hat{g}t}  H_{\hat{g}t} ,  k^\textrm{PFR}_{\hat{g}t} PFR_{\hat{g}t}$.   

Whereas nonlinearities such as the SOC for the nadir constraint are easily handled with off-the-shelf solvers, the bilinear terms will be addressed using binary expansion. In this case we choose to perform the binary expansion over the variables $P_{\hat{g}t}, H_{\hat{g}t}, PFR_{\hat{g}t}$ given that they are the common variables for the multiplications with $k^\textrm{E}_{\hat{g}t} , k^\textrm{H}_{\hat{g}t}, k^\textrm{PFR}_{\hat{g}t}$ and $\lambda^\textrm{E}_{t}, \lambda_{t}^{\textrm{H}}, \lambda^\textrm{PFR}_{t}$, which is described in the set of equations \eqref{eq:BE_x_strat}-\eqref{eq:kStratPFR_PFRStrat}. The number of intervals for the binary expansion for energy, inertia, and PFR provision are set as $\textrm{L}^{\textrm{E}}_{\hat{g}} = \textrm{L}^{\textrm{H}}_{\hat{g}}=\textrm{L}^{\textrm{PFR}}_{\hat{g}}=128$, which allows a good trade-off between solution accuracy and processing time. The length of the intervals are defined as $\Delta^{\textrm{E}}_{\hat{g}} = \frac{\textrm{N}_{\hat{g}}\textrm{P}^\textrm{max}_{\hat{g}}}{\textrm{L}^{\textrm{E}}_{\hat{g}}-1}$,  $\Delta^{\textrm{PFR}}_{\hat{g}} = \frac{\textrm{N}_{\hat{g}}\textrm{PFR}^\textrm{max}_{\hat{g}}}{\textrm{L}^{\textrm{PFR}}_{\hat{g}}-1}$, and $\Delta^{\textrm{H}}_{\hat{g}} = \frac{\textrm{N}_{\hat{g}} \overline{\textrm{H}}_{\hat{g}} \textrm{P}^\textrm{max}_{\hat{g}} }{\textrm{L}^{\textrm{H}}_{\hat{g}}-1}$, where $\textrm{N}_{\hat{g}}$ represents the 35 CCGTs units owned by the strategic player $\hat{g}$, representing a total installed capacity of 17.5GW as shown in Table~\ref{tab:GenerationMix}.

\vspace*{-3mm}

\begin{figure}[h]
    \centering
    \includegraphics{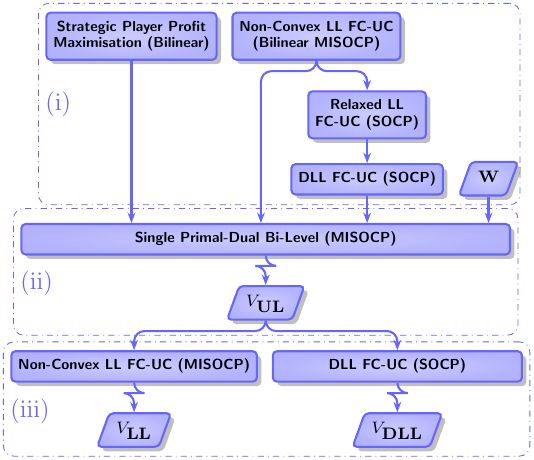}
    \caption{Solution methodology diagram.}
    \label{fig:Diagram}
    \vspace*{-4mm}
\end{figure}

\vspace*{-3mm}

\subsection{Solution Methodology} 
\label{ssec:W_choice}
Fig.~\ref{fig:Diagram} describes the methodology to solve the strategic bidding model proposed in this work. Block (i) describes the mathematical framework presented in sections \ref{sec:upperLevel}, \ref{sec:FreqSecUC}, \ref{sec:DualFreqSecUC}, representing the bilinear UL problem, the bilinear MISOCP primal LL problem, and the SOCP dual of the relaxed LL problem, respectively.

Block (ii) describes the Single Level Primal-Dual Bi-Level Formulation described in section \ref{ssec:singleLevel}, which represents a MISOCP, given the linearisation of the bilinear terms that appear in the objective function \eqref{eq:obj_singleLevel}, which are replaced by the expressions obtained in \eqref{eq:BE_x_strat}-\eqref{eq:kStratPFR_PFRStrat}. Note that block (ii) is a parametric model that depends on the value of $\textrm{W}$, as explained in section \ref{ssec:singleLevel}. As was also considered in \cite{Ye_Papadaskalopoulos_Kazempour}, block (ii) will only provide the strategic bidding multipliers $V_\textrm{UL}$ (i.e., $k^\textrm{E}_{\hat{g}t}, k^\textrm{H}_{\hat{g}t}, k^\textrm{PFR}_{\hat{g}t}$). In contrast, energy dispatch, AS provision, and their respective prices will be obtained in block (iii) for an accurate market-clearing solution.

Finally, block (iii) applies the strategic decisions $V_\textrm{UL}$ into the market-clearing problems (\ref{eq:LowerLevel}) and (\ref{eq:DualLowerLevel}) detailed in sections \ref{sec:FreqSecUC} and \ref{sec:DualFreqSecUC}, respectively. Block (iii) results allow the computation of the final quantities and prices to determine the profits of the strategic player for each penalty parameter W. 

Following \cite{Ye_Papadaskalopoulos_Kazempour}, the selection of $\textrm{W}$ is based on:

\begin{itemize}
    \item The accuracy of the Single Level solution in block (ii), determined by a small DG, and represented as $r^{DG}(\textrm{W})$. This value is calculated as the ratio between the DG in \eqref{eq:DG_singleLevel}, and the value of the LL system cost defined in eq. (\ref{eq:objFunc}), expressed as $r^{DG}(\textrm{W})=\frac{\eqref{eq:DG_singleLevel}}{\eqref{eq:objFunc}}$.
\end{itemize}

\begin{itemize}
    \item The strategic player profits level, $\Delta^{
    Profit}(\textrm{W})$, which is calculated in block (iii) as the difference between the strategic player profits (\ref{eq:objFunc_strat}) under strategic behaviour and its profits under competitive behaviour. This difference is normalised by the competitive profits and is expressed as $\Delta^{
    Profit}(\textrm{W})=\frac{(\ref{eq:objFunc_strat})^{\textrm{Strategic}} - (\ref{eq:objFunc_strat})^{\textrm{Competitive}}}{(\ref{eq:objFunc_strat})^{\textrm{Competitive}}}$.
\end{itemize}

\vspace*{-5mm}

\begin{alignat}{4} 
    & P_{\hat{g}t} = \sum_{n=0}^{\log_2(\textrm{L}^{\textrm{E}}_{\hat{g}})-1} 2^n  \Delta^{\textrm{E}}_{\hat{g}}  \; b_{\hat{g}tn}^{\textrm{E}} && \qquad \forall \hat{g}t && \label{eq:BE_x_strat} \\[3pt]
    & 0 \leq  \lambda^\textrm{E}_{t} - z_{\hat{g}tn}^{\lambda^\textrm{E}} \leq \textrm{M}^{\lambda^\textrm{E}}_{\hat{g}} (1-b_{\hat{g}tn}^{\textrm{E}})   && \qquad \forall \hat{g}tn && \label{eq:BE_lambdaE1} \\[3pt]
    & 0 \leq z_{\hat{g}tn}^{\lambda^\textrm{E}} \leq \textrm{M}^{\lambda^\textrm{E}}_{\hat{g}} \; b_{\hat{g}tn}^{\textrm{E}}   && \qquad \forall \hat{g}tn && \label{eq:BE_lambdaE2} \\[3pt]
    & \lambda^\textrm{E}_{t} P_{\hat{g}t} = \sum_{n=0}^{\log_2(\textrm{L}^{\textrm{E}}_{\hat{g}})-1} 2^n  \Delta^{\textrm{E}}_{\hat{g}}  \; z_{\hat{g}tn}^{\lambda^\textrm{E}} && \qquad \forall \hat{g}t && \label{eq:lambdaE_xStrat} \\[3pt]
    & 0 \leq k^\textrm{E}_{\hat{g}t} - z_{\hat{g}tn}^{{\textrm{k}^\textrm{E}}} \leq  \textrm{M}^{\textrm{k}^\textrm{E}}_{\hat{g}} (1-b_{\hat{g}tn}^{\textrm{E}})  && \qquad \forall \hat{g}t && \label{eq:BE_kStratE1} \\[3pt]
    & 0 \leq z_{\hat{g}tn}^{{\textrm{k}^\textrm{E}}} \leq  \textrm{M}^{\textrm{k}^\textrm{E}}_{\hat{g}} \; b_{\hat{g}tn}^{\textrm{E}}  && \qquad \forall \hat{g}t && \label{eq:BE_kStratE2} \\[3pt]
    &  k^\textrm{E}_{\hat{g}t}  P_{\hat{g}t}   =     \sum_{n=0}^{\log_2(\textrm{L}^{\textrm{E}}_{\hat{g}})-1} 2^n  \Delta^{\textrm{E}}_{\hat{g}} \;z_{\hat{g}tn}^{{\textrm{k}^\textrm{E}}} && \qquad \forall \hat{g}t && \label{eq:kStratE_xStrat}  \\[3pt]
%
    & H_{\hat{g}t} = \sum_{n=0}^{\log_2(\textrm{L}^{\textrm{H}}_{\hat{g}})-1} 2^n  \Delta^{\textrm{H}}_{\hat{g}} \; b_{\hat{g}tn}^{\textrm{H}} && \qquad \forall \hat{g}t && \label{eq:BE_H_strat} \\[3pt]
    & 0 \leq  \lambda^\textrm{H}_{t} - z_{\hat{g}tn}^{\lambda^\textrm{H}} \leq \textrm{M}^{\lambda^\textrm{H}}_{\hat{g}} (1-b_{\hat{g}tn}^{\textrm{H}})   && \qquad \forall \hat{g}tn && \label{eq:BE_lambdaH1} \\[3pt]
    & 0 \leq z_{\hat{g}tn}^{\lambda^\textrm{H}} \leq \textrm{M}^{\lambda^\textrm{H}}_{\hat{g}} \; b_{\hat{g}tn}^{\textrm{H}}   && \qquad \forall \hat{g}tn && \label{eq:BE_lambdaH2} \\[3pt]
    & \lambda^\textrm{H}_{t} H_{\hat{g}t} = \sum_{n=0}^{\log_2(\textrm{L}^{\textrm{H}}_{\hat{g}})-1} 2^n  \Delta^{\textrm{H}}_{\hat{g}}  \; z_{\hat{g}tn}^{\lambda^\textrm{H}} && \qquad \forall \hat{g}t && \label{eq:lambdaH_HStrat} \\[3pt]
    & 0 \leq k^\textrm{H}_{\hat{g}t} - z_{\hat{g}tn}^{{\textrm{k}^\textrm{H}}} \leq  \textrm{M}^{\textrm{k}^\textrm{H}}_{\hat{g}} (1-b_{\hat{g}tn}^{\textrm{H}})  && \qquad \forall \hat{g}t && \label{eq:BE_kStratH1} \\[3pt]
    & 0 \leq z_{\hat{g}tn}^{{\textrm{k}^\textrm{H}}} \leq  \textrm{M}^{\textrm{k}^\textrm{H}}_{\hat{g}} \; b_{\hat{g}tn}^{\textrm{H}}  && \qquad \forall \hat{g}t && \label{eq:BE_kStratH2} \\[3pt]
    & k^\textrm{H}_{\hat{g}t}  H_{\hat{g}t}   =     \sum_{n=0}^{\log_2(\textrm{L}^{\textrm{H}}_{\hat{g}})-1} 2^n  \Delta^{\textrm{H}}_{\hat{g}} \;z_{\hat{g}tn}^{{\textrm{k}^\textrm{H}}} && \qquad \forall \hat{g}t && \label{eq:kStratH_HStrat}  \\[3pt]
%
    & PFR_{\hat{g}t} = \sum_{n=0}^{\log_2(\textrm{L}^{\textrm{PFR}}_{\hat{g}})-1} 2^n  \Delta^{\textrm{PFR}}_{\hat{g}} \; b_{\hat{g}tn}^{\textrm{PFR}} && \qquad \forall \hat{g}t && \label{eq:BE_PFRl_strat} \\[3pt]
    & 0 \leq  \lambda^\textrm{PFR}_{t} - z_{\hat{g}tn}^{\lambda^\textrm{PFR}} \leq \textrm{M}^{\lambda^\textrm{PFR}}_{\hat{g}} (1-b_{\hat{g}tn}^{\textrm{PFR}})   && \qquad \forall \hat{g}tn && \label{eq:BE_lambdaPFR1} \\[3pt]
    & 0 \leq z_{\hat{g}tn}^{\lambda^\textrm{PFR}} \leq \textrm{M}^{\lambda^\textrm{PFR}}_{\hat{g}} \; b_{\hat{g}tn}^{\textrm{PFR}}   && \qquad \forall \hat{g}tn && \label{eq:BE_lambdaPFR2} \\[3pt]
    & \lambda^\textrm{PFR}_{t} PFR_{\hat{g}t} = \sum_{n=0}^{\log_2(\textrm{L}^{\textrm{PFR}}_{\hat{g}})-1} 2^n  \Delta^{\textrm{PFR}}_{\hat{g}}  \; z_{\hat{g}tn}^{\lambda^\textrm{PFR}} && \qquad \forall \hat{g}t && \label{eq:lambdaPFR_PFRStrat} \\[3pt]
    & 0 \leq k^\textrm{PFR}_{\hat{g}t} - z_{\hat{g}tn}^{{\textrm{k}^\textrm{PFR}}} \leq  \textrm{M}^{\textrm{k}^\textrm{PFR}}_{\hat{g}} (1-b_{\hat{g}tn}^{\textrm{PFR}})  && \qquad \forall \hat{g}t && \label{eq:BE_kStratPFR1} \\[3pt]
    & 0 \leq z_{\hat{g}tn}^{{\textrm{k}^\textrm{PFR}}} \leq  \textrm{M}^{\textrm{k}^\textrm{PFR}}_{\hat{g}} \; b_{\hat{g}tn}^{\textrm{PFR}}  && \qquad \forall \hat{g}t && \label{eq:BE_kStratPFR2} \\[3pt]
    & k^\textrm{PFR}_{\hat{g}t}  PFR_{\hat{g}t}   =     \sum_{n=0}^{\log_2(\textrm{L}^{\textrm{PFR}}_{\hat{g}})-1} 2^n  \Delta^{\textrm{PFR}}_{\hat{g}} \;z_{\hat{g}tn}^{{\textrm{k}^\textrm{PFR}}} && \qquad \forall \hat{g}t && \label{eq:kStratPFR_PFRStrat} 
\end{alignat}

\end{subequations}    

\vspace*{-4mm}

\begin{table*}
    \captionsetup{justification=centering, textfont={sc,footnotesize}, labelfont=footnotesize, labelsep=newline}
    \centering
    \renewcommand{\arraystretch}{1.2}
    \caption{Characteristics of power generators, RES and storage}
    \label{tab:GenerationMix}
    \begin{tabular}{p{3cm}|p{0.9cm}|p{0.9cm}|>{\columncolor{gray!10}}p{1.17cm}|p{1.17cm}|p{1.17cm}| p{0.9cm}|p{0.9cm}| p{0.9cm}| p{1.17cm}| p{1.45cm}} %
                               & Big Nuclear     & Nuclear     & Strategic CCGT  & CCGT & OCGT   & Offshore Wind   & Onshore Wind  & Solar PV   & PHES       & BESS\\ \hline
    Installed capacity (GW)     & 1.8            & 2.7        & 17.5  & 3.5   & 4     & 50.4        & 30      & 41      & 4.8     & 20\\ \hline
    Power range (MW)  ($\textrm{P}^\textrm{msg},  \textrm{P}^\textrm{max}$)   &1800, 1800  &1350, 1350  &250,500  &250,500   &50,100  &0,1800    &0,600  &0,250    &0,400    &0,50\\ \hline
    Inertia constant (s)  &4  &4   &5  &5 &5 &N/A  &N/A    &N/A   &2   &N/A \\ \hline
    Response type   &N/A      &N/A     &PFR &PFR &PFR  &N/A  &N/A  &N/A    &PFR   &EFR\\ \hline
    Response capacity (MW)  &N/A &N/A &$5\% \cdot \textrm{P}^\textrm{max}_g$  &$5\% \cdot \textrm{P}^\textrm{max}_g$  &$20\% \cdot \textrm{P}^\textrm{max}_g$  &N/A &N/A &N/A &$5\% \cdot \textrm{P}^\textrm{max}_s$  &$2.5\% \cdot \textrm{P}^\textrm{max}_s$\\ \hline
    Energy price offer $\textrm{O}^\textrm{E}_g$,\hspace{-0.5mm} $\textrm{O}^\textrm{E}_r$,\hspace{-0.5mm} $\textrm{O}^\textrm{E}_s$ (\pounds/MWh)  &10 &10  &130  &130  &200   &41  &36  &30   &46  &50\\ \hline
    Inertia price offer $\textrm{O}^\textrm{H}_g$,\hspace{-0.5mm} $\textrm{O}^\textrm{H}_s$ (\pounds/MWs)    &1 &1  &1.8  &1.8  &6  &N/A  &N/A  &N/A  &2  &N/A\\ \hline
    FR price offer $\textrm{O}^\textrm{PFR}_g$,\hspace{-0.5mm} $\textrm{O}^\textrm{PFR}_s$,\hspace{-0.5mm} $\textrm{O}^\textrm{EFR}_s$ (\pounds/MW)   &N/A &N/A  &15  &15  &50 &N/A  &N/A  &N/A  &30  &150\\ \hline
    \end{tabular}
\vspace*{-4mm}
\end{table*}

\section{Case Study} \label{sec:Results}

\begin{figure}[h]
    \centering
\includegraphics[width=0.975\linewidth]{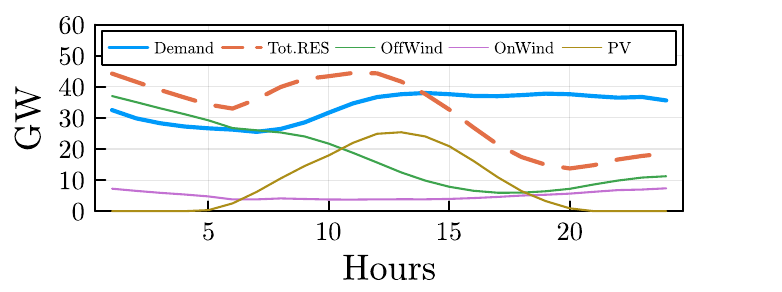}
    \vspace*{-1mm}
    \caption{Demand and RES availability profiles.}
\label{fig:input_profiles}
    \vspace*{-4mm}
\end{figure}

This case study is based on the `Leading the Way' scenario within National Grid's Future Energy Scenarios for 2030 \cite{NationalGridFES2023}, which provides an ambitious pathway towards decarbonisation in GB. National demand follows a typical profile for the GB system considering daily trends, while RES hourly profiles are obtained from \cite{PFENNINGER20161251, STAFFELL20161224}, as shown in Fig.~\ref{fig:input_profiles}. Generation technical parameters are included in Table~\ref{tab:GenerationMix}. Energy price offers represent fuel plus variable O\&M costs for thermal generators and storage and levelised cost of electricity for RES as published on \cite{beis2023electricity}. Inertia price offers are based on no-load plus start-up costs, i.e., the cost of being synchronised. FR price offers are based on National Grid's FR market information \cite{FRReport2022}. 
Frequency limits are based on GB regulation, considering $\textrm{RoCoF}_\textrm{max}=1\textrm{Hz/s}$ and $\Delta \textrm{f}_{\textrm{max}}=0.8\textrm{Hz}$. Response services, EFR and PFR, must be fully delivered in $\textrm{T}_\textrm{EFR}=1\textrm{s}$ and $\textrm{T}_\textrm{PFR}=10\textrm{s}$, respectively.

For sections \ref{ssec:strat_energy}, \ref{ssec:strat_AS}, \ref{ssec:strat_coupled}, the routine considers the methodology defined in section \ref{ssec:W_choice} for four different values of $\textrm{W}$, including $\textrm{W}=\{1, 10, 100, 1000\}$. In each section, the presented results correspond to the best $\textrm{W}$ solution, which yields the highest value of $\Delta^{Profit}(\textrm{W})$, that considers $r^{DG}(\textrm{W}) \leq 3\%$. 
The optimisation model is formulated in Julia JuMP \cite{Lubin2023} and the solver used is Gurobi \cite{gurobi}.

\vspace*{-3mm}

\vspace*{-3mm}

\subsection{Competitive Behaviour in the Energy and the Frequency-Containment Ancillary Services Markets} \label{ssec:competitive}

This section discusses the market-clearing in a perfectly competitive energy and AS market. 

\begin{figure}[h]
    \centering
    \includegraphics[width=0.975\linewidth]{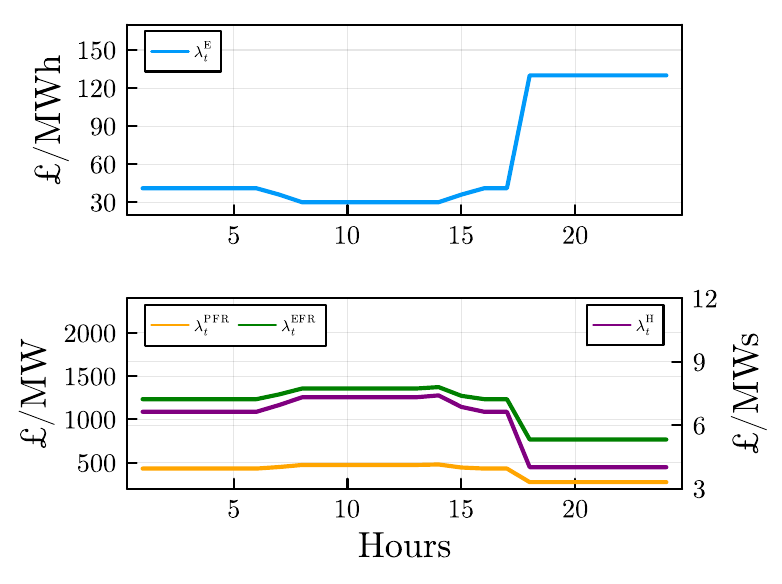}
    \vspace*{-3mm}
    \caption{Energy prices (top) and AS prices (bottom) under competitive behaviour in the energy and AS markets.  }
    \label{fig:compet_prices}
    \vspace*{-4mm}
\end{figure}

As shown in Fig.~\ref{fig:compet_prices}, the initial part of the day shows low energy and high AS prices. This pattern is due to the combination of high RES and low demand conditions until hour 17:00 (as seen in Fig.~\ref{fig:input_profiles}), significantly reducing the need for thermal generation to produce energy and bringing synchronous units online mainly to provide frequency security. The opposite trend occurs after hour 17:00 when demand is high, and RES is low. This pattern increases energy prices due to the need to provide energy with thermal units while decreasing AS prices due to the higher levels of inertia in the system.

In this case, the daily profits for the Strategic CCGT player, behaving competitively, are £3.326 million. This value will serve as a baseline for comparison with profit levels under strategic behaviour, $\Delta^{Profit}(\textrm{W})$, as presented in Table \ref{tab:ResultsW} and discussed in the following sections.


\vspace*{-1mm}

\begin{table}[h]
    \captionsetup{justification=centering, textfont={sc,footnotesize}, labelfont=footnotesize, labelsep=newline}
    \centering
    \renewcommand{\arraystretch}{1.2}
    \caption{Results}
    \label{tab:ResultsW}
    \begin{tabular}{p{1.4cm}|>{\centering\arraybackslash}p{0.5cm}|>{\centering\arraybackslash}p{1cm}|>{\centering\arraybackslash}p{1.4cm}|>{\centering\arraybackslash}p{1.3cm}} 
Strategic Behaviour & \textrm{W}   & $r^{DG}(\textrm{W})$  & $\Delta^{Profit}(\textrm{W})$ & Computation Time (s) \\   \hline
\multirow{4}{1.4cm}{In the energy market}         & 1    & 7.80\% & 71.14\%      & 28.79  \\ 
                                                  & \cellcolor{gray!10}10   & \cellcolor{gray!10}2.57\% & \cellcolor{gray!10}16.59\%       & \cellcolor{gray!10}9.54   \\ 
                                                  & 100  & 0.22\% & 16.40\%       & 40.91  \\ 
                                                  & 1000 & 0.22\% & 8.34\%       & 834.82 \\ \hline
\multirow{4}{1.4cm}{In the AS market}             & 1    & 9.30\% & 102.82\%      & 32.62  \\ 
                                                  & \cellcolor{gray!10}10   & \cellcolor{gray!10}2.41\% & \cellcolor{gray!10}126.31\%     & \cellcolor{gray!10}8.33   \\ 
                                                  & 100  & 0.22\% & -2.18\%      & 8.20   \\ 
                                                  & 1000 & 0.22\% & -0.27\%       & 9.72   \\ \hline
\multirow{4}{1.4cm}{In the energy and AS markets} & 1    & 7.60\% & 222.88\%     & 29.60  \\ 
                                                  & \cellcolor{gray!10}10   & \cellcolor{gray!10}2.36\% & \cellcolor{gray!10}165.06\%     & \cellcolor{gray!10}10.90  \\ 
                                                  & 100  & 0.21\% & 16.17\%       & 41.98  \\ 
                                                  & 1000 & 0.21\% & 8.34\%       & 1300.88  \\ \hline
\end{tabular}
\vspace*{-4mm}
\end{table}

\subsection{Strategic Behaviour in the Energy Market} \label{ssec:strat_energy}
This section analyses the well-known case of strategic behaviour in the energy market. To make a fair comparison with the following cases, we consider the requirement and provision of frequency-containment AS but we assume a competitive behaviour in the AS market (i.e., $ \overline{\textrm{k}}^\textrm{H}_{\hat{g}}= \overline{\textrm{k}}^\textrm{PFR}_{\hat{g}}=1$). Thus, the strategic behaviour only depends on the energy bidding of the Strategic CCGT, represented by the multiplier $k^\textrm{E}_{\hat{g}t}$, limited by an upper limit $ \overline{\textrm{k}}^\textrm{E}_{\hat{g}} = 3$. 

As shown in Table \ref{tab:ResultsW}, the best solution is obtained for $\textrm{W}=10$, which presents a small DG ratio, $r^{DG}(\textrm{W})=2.57\%$, and a profit increase of $\Delta^{Profit}(\textrm{W})=16.59\%$ regarding the competitive case presented in section \ref{ssec:competitive}, reaching total daily profits of £3.877 million. 

\begin{figure}[t!]
    \centering
    \includegraphics[width=1\linewidth]{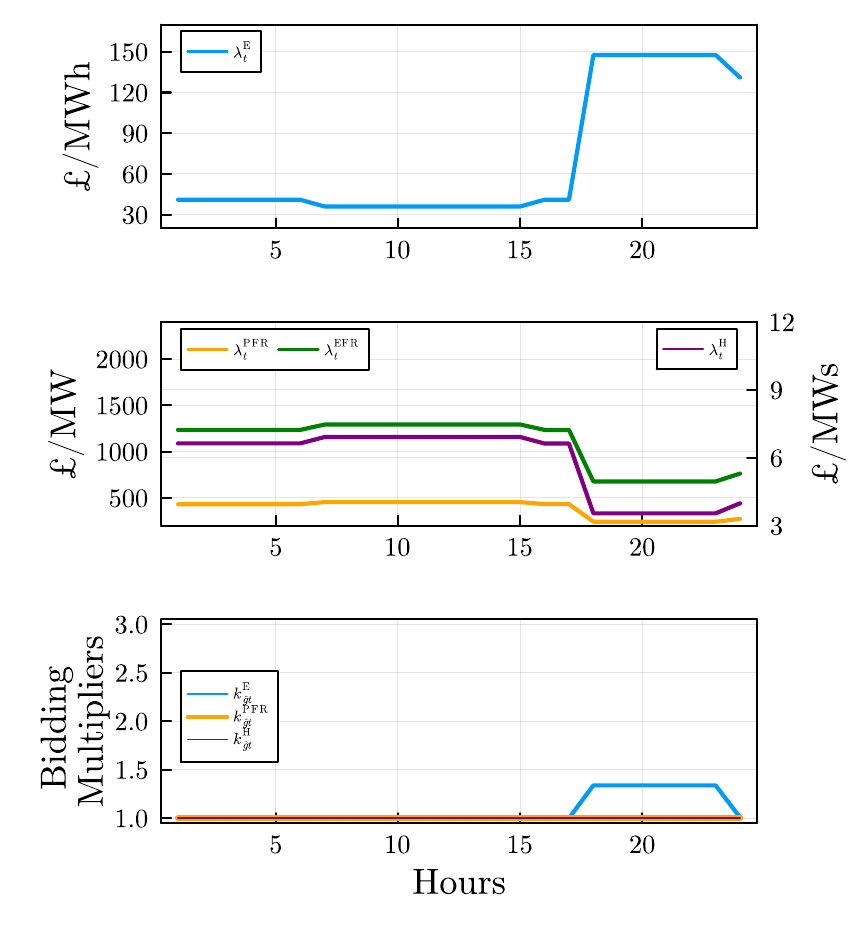}
    \vspace*{-9mm}
    \caption{Energy prices (top), AS prices (middle), and bidding multipliers (bottom) under strategic behaviour in the energy market. } 
    \label{fig:stratE_prices}
    \vspace*{-8mm}
\end{figure}

Regarding resulting prices, as can be seen in Fig. \ref{fig:stratE_prices}, there are no relevant changes in the AS market. At the same time, there is a slight increase in energy prices after hour 17:00. This is due to the strategic energy bidding multipliers, which reach a value of $k^\textrm{E}_{\hat{g}t}=1.34$. Even though the maximum limit of the energy bidding multiplier is set at $ \overline{\textrm{k}}^\textrm{E}_{\hat{g}} = 3$, there is no relevant strategic game that can be played in this market. This is due to the variety of energy sources in the GB system for 2030. As energy is widely available, there is less room for increasing profits through behaving strategically in the energy market.

\subsection{Strategic Behaviour in the Frequency-Containment Ancillary Services Market} \label{ssec:strat_AS}
In this section, we analyse strategic behaviour in the frequency-containment AS market. To do this, we consider a competitive behaviour in the energy market ($\overline{\textrm{k}}^\textrm{E}_{\hat{g}}=1$) while allowing strategic bidding offers in the inertia and PFR markets, considering as multipliers limits $\overline{\textrm{k}}^\textrm{H}_{\hat{g}t}=\overline{\textrm{k}}^\textrm{PFR}_{\hat{g}t}=3$.  

Table~\ref{tab:ResultsW} shows a substantial increase in profits when the Strategic CCGT behaves strategically in the AS market. This result is obtained with $\textrm{W}=10$ which has $r^{DG}(\textrm{W})=2.41\%$ and an increase of $\Delta^{Profit}(\textrm{W})=126.31\%$, accounting for a total of £7.526 million profits, which double the profits obtained with competitive behaviour in section \ref{ssec:competitive}.

\begin{figure}
    \centering
    \includegraphics[width=1\linewidth]{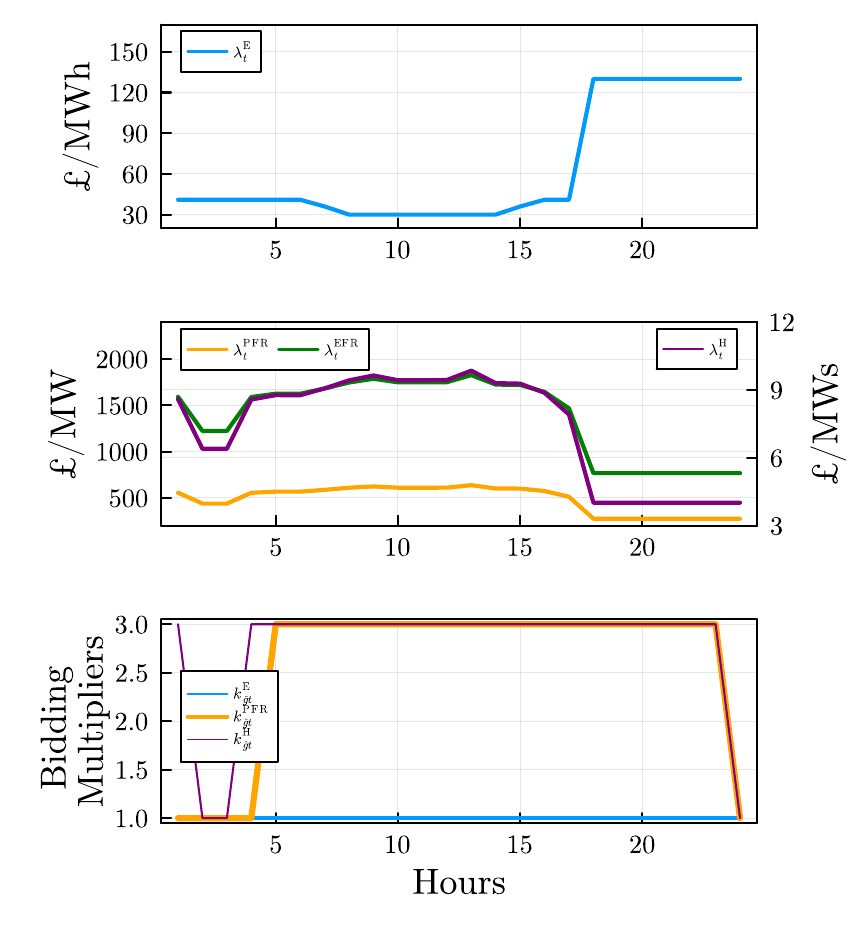}
    \vspace*{-9mm}
    \caption{Energy prices (top), AS prices (middle), and bidding multipliers (bottom) under strategic behaviour in the AS market. } 
    \label{fig:stratAS_prices}
    \vspace*{-7mm}
\end{figure}

The strategic bidding can be seen in Fig.~\ref{fig:stratAS_prices}. In this case, the strategic player increases its inertia, and PFR bidding offers prices in most of the hours of the day, reaching the bidding limits with $k^\textrm{H}_{\hat{g}t}=3$ and $k^\textrm{PFR}_{\hat{g}t}=3$. As there are fewer AS providers in most hours of the day, there is more room to change AS prices compared with the strategic behaviour in the energy market, as described in section \ref{ssec:strat_energy}.

\subsection{Strategic Behaviour in the Energy and the Frequency-Containment Ancillary Services Markets} \label{ssec:strat_coupled}
This section analyses the combined effect of strategic behaviour presented in sections \ref{ssec:strat_energy} and \ref{ssec:strat_AS}. Thus, the Strategic CCGT can simultaneously induce market power in the energy and the AS markets (i.e., $\overline{\textrm{k}}^\textrm{E}_{\hat{g}t}=\overline{\textrm{k}}^\textrm{H}_{\hat{g}t}=\overline{\textrm{k}}^\textrm{PFR}_{\hat{g}t}=3$).

As can be seen in Table~\ref{tab:ResultsW}, the best solution is obtained for $\textrm{W}=10$, with $r^{DG}(\textrm{W})=2.36\%$, and $\Delta^{Profit}(\textrm{W})=165.06\%$, reaching total daily profits of £8.815 millions. 

\begin{figure}[t!]
    \centering
    \includegraphics[width=1\linewidth]{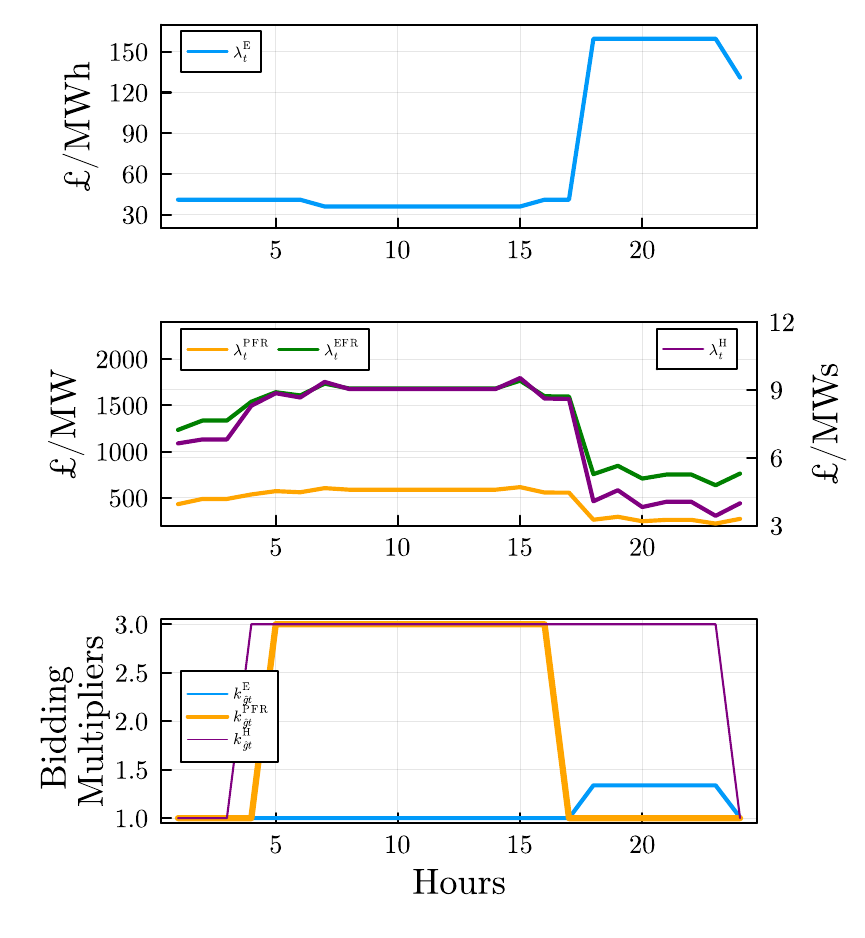}
    \vspace*{-9mm}
    \caption{Energy prices (top), AS prices (middle), and bidding multipliers (bottom) under strategic behaviour in the energy and AS markets. } 
    \label{fig:stratEAS_prices}
    \vspace*{-7mm}
\end{figure}

The bidding offer prices are illustrated in Fig.~\ref{fig:stratEAS_prices}, where it can be seen that the strategy combines elements from the two strategies presented in  \ref{ssec:strat_energy} and \ref{ssec:strat_AS}, keeping high AS bidding offers during most of the hours of the day, while increasing energy bidding offers only in hours with high net demand.

To evaluate the impact of strategic behaviour in the energy market, it is noteworthy that daily profits, in this case, are 17.13\% higher than the isolated strategic behaviour in the AS market presented in section \ref{ssec:strat_AS}. This increase is similar to the 16.57\% rise in profits observed when comparing isolated strategic behaviour in the energy market (section \ref{ssec:strat_energy}) with the competitive behaviour (section \ref{ssec:competitive}).

The significant impact of strategic behaviour in the AS market becomes evident when noticing the daily profits increase of 127.37\% of this case compared with the isolated strategic behaviour in the energy market (section \ref{ssec:strat_energy}). Similarly, the increase in profits when considering the isolated strategic behaviour in the AS market (section \ref{ssec:strat_AS}) compared with competitive behaviour (section \ref{ssec:competitive}) represented a rise of 126.28\%.

Thus, while the strategic player’s influence in the energy market increases its profits by $\sim$17\%, the strategic behaviour in the AS market can double its profits, highlighting the need for regulation to prevent market power in the AS market.

\vspace*{-3mm}

\section{Conclusion} \label{sec:Conclusion}
This work developed a framework for analysing strategic bidding in the energy and frequency-containment AS markets. The case study focused on the GB system in 2030, for which we considered a strategic player, a firm based on CCGT units with a substantial presence in the electricity market. As demonstrated, this player's ability to influence energy and AS prices can have detrimental effects, particularly in the AS market, highlighting the importance of effective market regulation.

Future work would involve modelling the strategic behaviour of multiple market players, including RES, BESS units, electrolyzers, and interconnectors, by formulating an Equilibrium Problem with Equilibrium Constraints.

\vspace*{-5mm}

\appendices
\ifCLASSOPTIONcaptionsoff
  \newpage
\fi

\bibliographystyle{IEEEtran} 
\bibliography{Carlos_PhD}
\end{document}